\documentclass[12pt]{iopart}

\usepackage{iopams}
\usepackage{setstack}
\usepackage{graphicx}
\usepackage{pict2e}
\usepackage{color}
\usepackage{hyperref}

\newcommand{\C}[2]{{#1 \choose #2}}
\renewcommand{\O}[1]{\mathcal{O}\left(#1\right)}

\newcommand{\Li}{\mathrm{Li}}
\newcommand{\sgn}{\mathrm{sgn}}
\renewcommand{\Re}{\mathrm{Re}}
\renewcommand{\Im}{\mathrm{Im}}
\newcommand{\openone}{\mbox{{\small 1}$\!\!$1}}

\newcommand{\affiliation}[1]{\address{#1}}
\renewcommand{\pacs}[1]{\noindent\textbf{PACS numbers:} #1}
\newcommand{\keywords}[1]{\noindent\textbf{Keywords:} #1}
\newcommand{\tfrac}[2]{\mbox{\small$\frac{#1}{#2}$}}
\renewcommand{\text}[1]{\mathrm{#1}}

\begin{document}

\title[Spectrum of periodic TASEP - first excited states]{Spectrum of the totally asymmetric simple exclusion process on a periodic lattice - first excited states}
\author{Sylvain Prolhac}
\affiliation{Laboratoire de Physique Th\'eorique, IRSAMC, UPS, Universit\'e de Toulouse, France\\Laboratoire de Physique Th\'eorique, UMR 5152, Toulouse, CNRS, France}
\date{\today}

\begin{abstract}
We consider the spectrum of the totally asymmetric simple exclusion process on a periodic lattice of $L$ sites. The first eigenstates have an eigenvalue with real part scaling as $L^{-3/2}$ for large $L$ with finite density of particles. Bethe ansatz shows that these eigenstates are characterized by four finite sets of positive half-integers, or equivalently by two integer partitions. Each corresponding eigenvalue is found to be equal to the value at its saddle point of a function indexed by the four sets. Our derivation of the large $L$ asymptotics relies on a version of the Euler-Maclaurin formula with square root singularities at both ends of the summation range.\\

\\
%02.30.Ik Integrable systems
%02.50.Ga Markov processes
%05.40.-a Fluctuation phenomena, random processes, noise, and Brownian motion
%05.60.Cd Classical transport

\end{abstract}

%\today
\maketitle
%\tableofcontents\newpage

%%%%%%%%%%%%%%%%%
%%             %%
%%  Section I  %%
%%             %%
%%%%%%%%%%%%%%%%%
\begin{section}{Introduction}
\label{section introduction}
The one dimensional totally asymmetric simple exclusion process (TASEP) \cite{D1998.1,S2001.1,GM2006.1,S2007.1,M2011.1} is a Markov process featuring classical hard-core particles moving in the same direction on a lattice. It has been much studied in the last 20 years as a simple example of a genuine non-equilibrium process with transition rates breaking detailed balance. Another interest in TASEP, as well as its generalization ASEP where particles hop in both direction with different rates, comes from a mapping to a microscopic model of a growing interface. At large scales, the dynamics of the interface is described by a very singular \cite{H2014.1} nonlinear stochastic partial differential equation called the Kardar-Parisi-Zhang (KPZ) equation \cite{KPZ1986.1,HHZ1995.1,SS2010.4,KK2010.1,C2011.1}.

Models in the one dimensional KPZ universality class are characterized by a dynamical exponent $z=3/2$, which means that the relaxation time grows as $L^{3/2}$ for large system size $L$. The relaxation time is equal to the inverse of the gap between the stationary eigenvalue of the time evolution operator and the eigenvalue of the first excited state. The large $L$ asymptotics of the gap has been computed exactly for periodic TASEP \cite{D1987.1,GS1992.1,GS1992.2,GM2004.1,GM2005.1} and ASEP \cite{K1995.1}, as well as for TASEP \cite{dGE2005.1,dGE2006.1} and ASEP \cite{dGE2008.1,dGFS2011.1} on an open interval connected to reservoirs of particles, and also for generalizations of ASEP with several species of particles \cite{AKSS2009.1,WK2010.1} and with extended particles \cite{AB1999.1}. The gap of more general reaction diffusion processes has also been studied \cite{H2003.1}.

Many exact results have also been obtained for the fluctuations and large deviations \cite{D2007.1,T2009.1} of exclusion processes observed at large scales. In the stationary state, with a time of observation much larger than the relaxation time, explicit formulas have been derived for the large deviation function of the current of particles for periodic TASEP \cite{DL1998.1,DA1999.1} and ASEP \cite{LK1999.1,ADLvW2008.1,PM2009.1,P2010.1,PSS2010.1,S2011.1}, and also for TASEP \cite{LM2011.1} and ASEP \cite{dGE2011.1,GLMV2012.1,L2013.1,LP2014.1} with open boundaries or with several species of particles \cite{DE1999.1,C2008.1}. These large deviation functions, which are in general non-Gaussian, verify a Gallavotti-Cohen symmetry relation \cite{GC1995.1,BCHM2010.1}. On the other hand, the transient regime, with times of observation much smaller than the relaxation time, has been studied a lot in the context of height fluctuations for KPZ universality class \cite{PS2000.1,S2006.1,FS2011.1,TSSS2011.1}. The probability distribution of the current fluctuations in TASEP \cite{J2000.1,PS2002.1} and ASEP \cite{TW2009.1,SS2010.1,ACQ2011.1} have been found to be related to known distributions from random matrix theory, see also \cite{D2010.1,CLDR2010.1} for related results for the problem of a directed polymer in a random medium, which also belong to KPZ universality class.

A clear link is still missing between the results for the current fluctuations obtained in the stationary and in the transient regime, see however \cite{LK2006.1,BPPP2006.1,GMGB2007.1,PBE2011.1} for studies in the crossover regime, with an observation time of the order of the relaxation time. An exact calculation of the crossover behaviour would presumably require a summation over the corresponding eigenstates, and thus the knowledge of all eigenvectors and eigenvalues. This motivates further study of the spectrum beyond the gap. In \cite{P2013.1}, the bulk of the spectrum of TASEP was studied, with eigenvalues scaling proportionally to $L$ for large system size $L$. We consider here the first eigenvalues, which scale as $L^{-3/2}$ like the gap.

Explicit calculations for exclusion processes are made possible by the use of Bethe ansatz, which was initially introduced to diagonalize the Hamiltonian of Heisenberg spin chain. Exclusion processes possess the same kind of mathematical structures as quantum integrable models, although they are models of classical particles. In particular, one can define \cite{GM2006.1} for them a family of transfer matrices which commute with each other as a consequence of Yang-Baxter equation. Several aspects of the integrability of exclusion processes have been studied, for periodic systems \cite{P2003.1,GM2006.2,GM2007.1,GM2007.2}, with open boundaries \cite{S2009.1,CRS2010.1,CRS2011.1}, on the infinite line \cite{S1997.1,PP2006.1,TW2008.1}, and with more general hopping rules \cite{AKK1999.1}.

We study in this paper the structure of eigenvalues of periodic TASEP for higher excited states beyond the gap. We use Bethe ansatz in section \ref{section low spectrum} to write the eigenvalues for finite number of particles on a finite lattice. The large $L$ asymptotics performed in section \ref{section asymptotic expansions} relies on a version of the Euler-Maclaurin formula with square root singularities at both ends of the summation range (\ref{S[zeta]}), which is somewhat simpler than the previous derivations in the case of the gap. Section \ref{section eta} is devoted to the study of a function in terms of which the eigenvalues are expressed. A numerical study of the first eigenvalues is made in section \ref{section numerics}, as well as some asymptotics when one goes deeper into the spectrum.
\end{section}

%%%%%%%%%%%%%%%%%
%%             %%
%%  Section 2  %%
%%             %%
%%%%%%%%%%%%%%%%%
\begin{section}{First excited states}
\label{section low spectrum}

\begin{subsection}{The Markov matrix}
We consider the one-dimensional TASEP on a periodic lattice with $L$ sites. The $N$ particles of the system, $0<N<L$, move in continuous time by hopping from site to site in the forward direction. A particle can only hop if the destination site is empty, so that at any time each site is either empty or occupied by exactly one particle. Particles hop at a distance $1$ on the lattice, with hopping rate equal to $1$: in a small time interval $dt$, each particle moves with probability $dt$ if the site next to it in the forward direction is empty. The definition of the model is summarized in figure \ref{fig TASEP}.

The time evolution is encoded in a Markov matrix $M$, whose non-diagonal entries encode the transition rates between configurations: for $\mathcal{C}\neq\mathcal{C}'$, the matrix element $M(\mathcal{C}',\mathcal{C})\geq0$ is the rate at which the system goes from configuration $\mathcal{C}$ to configuration $\mathcal{C}'$. This rate is either $1$ if $\mathcal{C}'$ can be obtained from $\mathcal{C}$ by moving one and only one particle, and $0$ otherwise. The diagonal entries $M(\mathcal{C},\mathcal{C})$ are equal to minus the total exit rate from configuration $\mathcal{C}$.

Since the dynamics of TASEP breaks detailed balance, the eigenvalues $E$ of $M$ are in general complex numbers. They verify $\Re\,E<0$ except for the stationary state which has $E=0$. The relaxation toward the stationary state is governed by the eigenstate whose eigenvalue has the smallest real part, stationary eigenvalue excepted. This minimal value of $|\Re\,E|$ is called the gap, and is equal to the inverse of the relaxation time. It was shown in \cite{D1987.1,GS1992.1,GS1992.2,GM2004.1,GM2005.1} that in the thermodynamic limit $L,N\to\infty$ with density $\rho=N/L$ fixed, the gap scales as $L^{-3/2}$.
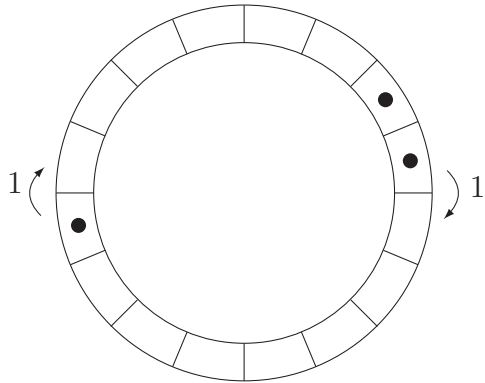
\begin{figure}
  \begin{center}
    \begin{picture}(70,50)
      \put(3.5,25){$1$}
      \put(65,24.5){$1$}
      \put(35,25){\circle{40}}
      \put(35,25){\circle{50}}
      \qbezier(8,22)(5,25)(8,28)
      \put(35,45){\line(0,1){5}}
      \put(55,25){\line(1,0){5}}
      \put(35,5){\line(0,-1){5}}
      \put(13,20.7){\circle*{2}}
      \put(15,25){\line(-1,0){5}}
      \put(53.8,37.4){\circle*{2}}
      \put(57.1,29.3){\circle*{2}}
      \qbezier(62,28)(65,25)(62,22)
      \put(8,28){\vector(1,1){0.5}}
      \put(62,22){\vector(-1,-1){0.5}}
      \put(49.2,39.2){\line(1,1){3.54}}
      \put(49.2,10.8){\line(1,-1){3.54}}
      \put(20.8,39.2){\line(-1,1){3.54}}
      \put(42.6,43.48){\line(5,12){1.92}}
      \put(53.48,32.6){\line(12,5){4.62}}
      \put(20.8,10.8){\line(-1,-1){3.54}}
      \put(42.6,6.52){\line(5,-12){1.92}}
      \put(53.48,17.4){\line(12,-5){4.62}}
      \put(16.52,32.6){\line(-12,5){4.62}}
      \put(27.4,43.48){\line(-5,12){1.92}}
      \put(27.4,6.52){\line(-5,-12){1.92}}
      \put(16.52,17.4){\line(-12,-5){4.62}}
    \end{picture}
  \end{center}
  \caption{The totally asymmetric simple exclusion process (TASEP) on a one-dimensional lattice with periodic boundary conditions. Particles hop only in the clockwise direction.}
  \label{fig TASEP}
\end{figure}
\end{subsection}

\begin{subsection}{The transfer matrix}
\label{subsection transfer matrix}
The quantum integrability of TASEP can be understood in terms of a family $t(\lambda)$ of commuting transfer matrices, $t(\lambda)t(\mu)=t(\mu)t(\lambda)$ for any $\lambda$ and $\mu$, see \textit{e.g.} \cite{GM2006.1}. This commutation relation is a consequence of Yang-Baxter equation. The parameter $\lambda$ of the transfer matrix is usually called the spectral parameter.

For $\lambda\in[0,1]$, the transfer matrix is the Markov matrix of a process in discrete time with parallel update and long range hopping, with transition probabilities depending on $\lambda$ as a geometric progression with the total distance travelled \cite{GM2007.1}. We show in section \ref{subsection expansion E(lambda)} that the real part of the first eigenvalues of $t(\lambda)-\openone$ scale as $L^{-3/2}$ for $\lambda\in[0,1]$: the relaxation time of the model with long range hopping described by $t(\lambda)$ grows as $L^{3/2}$ similarly to TASEP.

The Markov matrix $M$ of TASEP is equal to $M=t'(0)t^{-1}(0)$. One can also define from the transfer matrix generalized "Hamiltonians" $F_{k}=[\partial_{\lambda}^{k}\log t(\lambda)]_{|\lambda\to0}$ which commute with $M$ and contain only interactions on at most $k+1$ neighbouring sites \cite{GM2007.1,GM2007.2}. The operators $F_{k}$, $k\geq2$ are in general not Markovian as some of their non-diagonal entries are negative.
\end{subsection}

\begin{subsection}{Eigenvalues from Bethe ansatz}
\label{subsection Bethe ansatz}
Each eigenvector $|\Psi\rangle$ of the Markov matrix $M$, and more generally of the transfer matrix $t(\lambda)$, can be expressed in terms of $N$ Bethe roots $y_{1}$, \ldots, $y_{N}$ (that depend on the eigenstate) as
\begin{equation}
\label{eigenvector}
\Psi(x_{1},\ldots,x_{N})=\langle x_{1},\ldots,x_{N}|\Psi\rangle=\det\Big[\Big(\frac{y_{k}}{y_{j}}\Big)^{N-j}(1-y_{k})^{x_{j}}\Big]_{j,k=1,\ldots,N}\;.
\end{equation}
The corresponding eigenvalue $E$ of $M$ is equal to
\begin{equation}
\label{E[y]}
E=\sum_{j=1}^{N}\frac{y_{j}}{1-y_{j}}\;,
\end{equation}
and the Bethe roots must be solution of the Bethe equations
\begin{equation}
\label{Bethe equations}
(1-y_{j})^{L}=(-1)^{N-1}\prod_{k=1}^{N}\frac{y_{j}}{y_{k}}\;.
\end{equation}
The stationary eigenstate corresponds to $y_{j}=0$ for all $j$, for which the right hand side of Bethe equations is not well defined. This solution can in fact be understood by the introduction of a small twist $\gamma>0$, see \textit{e.g.} \cite{P2013.1}, and by taking the limit $\gamma\to0$. Apart from this peculiarity, it is believed that (\ref{E[y]}), (\ref{Bethe equations}) gives all the other eigenvalues of $M$, although a rigorous proof is still missing (see however \cite{PP2007.1} for a discrete time case).

We define the parameter
\begin{equation}
\label{b[y]}
b=\frac{1}{L}\sum_{j=1}^{N}\log y_{j}\;.
\end{equation}
Taking the power $1/L$ of the Bethe equations (\ref{Bethe equations}), it follows that there must exist numbers $k_{j}$, integers if $N$ is odd, half-integers if $N$ is even, such that
\begin{equation}
\label{Bethe equations[b] 1/L}
\frac{1-y_{j}}{y_{j}^{\rho}}=\rme^{\frac{2\rmi\pi k_{j}}{L}-b}\;,
\end{equation}
where $\rho=N/L$ is the density of particles. By analogy with the case of undistinguishable free particles (see \textit{e.g.} \cite{P2013.1}, Appendix D), the numbers $k_{j}$ must be distinct modulo $L$. They can be ordered as $0<k_{1}<\ldots<k_{N}\leq L$ so that each eigenstate is counted only once.

It turns out that the function $g:y\mapsto(1-y)/y^{\rho}$ is a bijection of the complex plane (minus the cut $\mathbb{R}^{-}$ and its image by $g$, see \ref{appendix psi}). Eq (\ref{Bethe equations[b] 1/L}) can then be solved for $y_{j}$ in terms of the reciprocal function $g^{-1}$. In the following, we use the function $\psi=\log g^{-1}$ instead of $g^{-1}$. It can be defined as the solution of the implicit equation
\begin{equation}
\label{psi}
\rme^{\psi(z)}+z\,\rme^{\rho\psi(z)}=1\;
\end{equation}
such that $-\pi<\Im\,\psi(z)<\pi$. In particular, at half-filling $\rho=1/2$, one has $\psi(z)=-2\log(z/2+\sqrt{1+z^{2}/4})$, while $\psi(z)\to\log(1-z)$ for $\rho\to0$ and $\psi(z)\to-\log(1+z)$ for $\rho\to1$. Some properties of the function $\psi$ are discussed in \ref{appendix psi} for general $\rho$. A plot of contour lines of $\psi$ is given in figure \ref{fig psi contour}.

Writing
\begin{equation}
\label{y[k,b]}
y_{j}=\exp\Big(\psi\Big(\rme^{\frac{2\rmi\pi k_{j}}{L}-b}\Big)\Big)\;,
\end{equation}
we can finally express the eigenvalues of the Markov matrix in terms of the function $\psi$ as
\begin{equation}
\label{E[psi]}
E=\sum_{j=1}^{N}\Big(\exp\Big[-\psi\Big(\rme^{\frac{2\rmi\pi k_{j}}{L}-b}\Big)\Big]-1\Big)^{-1}\;,
\end{equation}
where the parameter $b$ is solution of
\begin{equation}
\label{b[psi]}
b=\frac{1}{L}\sum_{j=1}^{N}\psi\Big(\rme^{\frac{2\rmi\pi k_{j}}{L}-b}\Big)\;.
\end{equation}

The translation operator commutes with the Markov matrix. From (\ref{eigenvector}), its eigenvalue $\rme^{2\rmi\pi P/L}$, $P\in\mathbb{Z}$ for the eigenstate with Bethe roots $y_{j}$ is $\rme^{2\rmi\pi P/L}=\prod_{j=1}^{N}(1-y_{j})$. Using (\ref{Bethe equations[b] 1/L}) and (\ref{b[y]}), one has $\rme^{2\rmi\pi P/L}=\prod_{j=1}^{N}\rme^{2\rmi\pi k_{j}/L}$, thus
\begin{equation}
\label{P[k]}
P=\sum_{j=1}^{N}k_{j}
\quad
\text{mod}\,$L$\;.
\end{equation}
\begin{figure}
  \begin{center}
  \includegraphics[width=100mm]{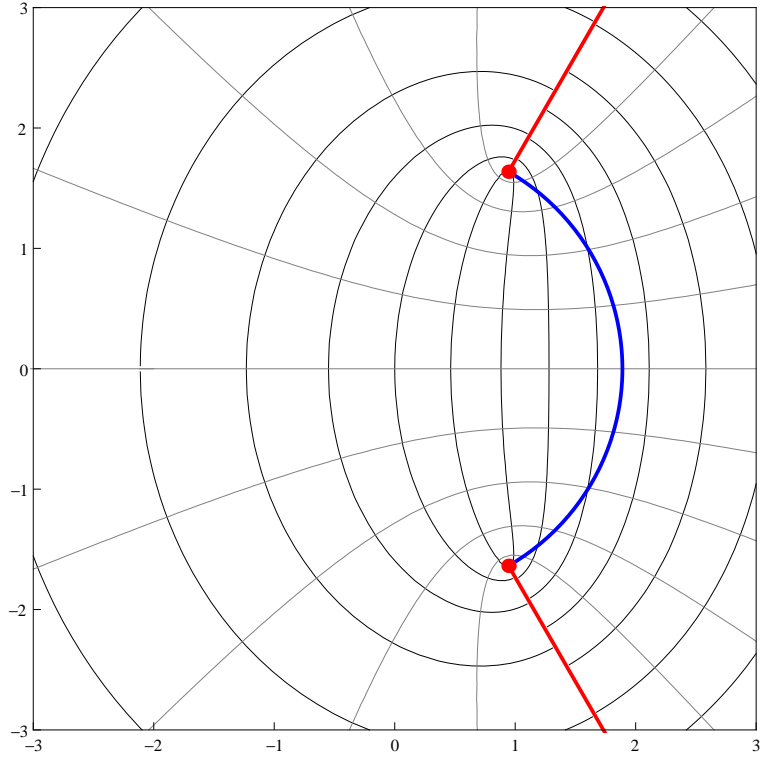}
  \end{center}
  \caption{Contour lines of the function $\psi$, defined implicitely in (\ref{psi}), for $\rho=1/3$. The black curves correspond to the $z$ of the complex plane such that $\Re\,\psi(z)=2,\,1.5,\,1,\,\ldots,\,-4$ from left to right. The gray curves correspond to the $z$ such that $\Im\,\psi(z)=-\frac{4\pi}{5},\,-\frac{3\pi}{5},\,\ldots,\,\frac{4\pi}{5}$ from up to down. The two thick, red lines are the branch cuts of  $\psi$. The thick, blue curve that links the two branch points is the set of the $z=\rme^{2\rmi\pi u-b_{0}}$ for $-\frac{\rho}{2}<u<\frac{\rho}{2}$ with $b_{0}$ defined in (\ref{b0}): this is the contour used in the integrals of equations (\ref{Euler-Maclaurin b}) and (\ref{Euler-Maclaurin E}).}
  \label{fig psi contour}
\end{figure}
\end{subsection}

\begin{subsection}{Lowest eigenvalues}
The stationary state $E=0$ corresponds in principle to the choice $k_{j}=k_{j}^{(0)}$, with
\begin{equation}
\label{k0}
k_{j}^{(0)}=j-\frac{N+1}{2}
\quad
\text{mod}\;L\;.
\end{equation}
Numerical studies seem to indicate that for this choice, (\ref{b[psi]}) does not have any finite solution for $b$. This problem can be avoided if one introduces a small twist, as mentioned in section \ref{subsection Bethe ansatz}. For all the other choices of the $k_{j}$'s, however, (\ref{b[psi]}) seems to have a unique solution for $b$.\\\indent
The first excitations above the stationary state (corresponding to eigenvalues with real part closest to $0$) correspond to sets of $k_{j}$'s obtained from $k_{j}^{(0)}$ by moving a finite number of $k_{j}^{(0)}$ of a finite distance for large $L$, with the density of particles $\rho=N/L$ taken fixed. Since the $k_{j}^{(0)}$'s completely fill the interval $[-\rho L/2,\rho L/2]$, one can only move $k_{j}^{(0)}$'s that are close to $\pm\rho L/2$. In particular, the first excited state (or gap) has a doubly degenerate real part: the corresponding eigenvalues $E_{1}^{\pm}$, which are complex conjugate of each other, correspond to $k_{j}=k_{j}^{(0)}$, $j=1,\ldots,N-1$, $k_{N}=(N+1)/2$ and to $k_{j}=k_{j}^{(0)}$, $j=2,\ldots,N$, $k_{1}=-(N+1)/2$, see figure \ref{fig choice k}.
\begin{figure}
  \begin{center}
    \begin{picture}(150,46)
      \put(0,40){\color[rgb]{0.7,0.7,0.7}\polygon*(40,0)(110,0)(110,5)(40,5)}
      \multiput(0,40)(5,0){30}{\polygon(0,0)(5,0)(5,5)(0,5)}
      \put(0,30){\color[rgb]{0.7,0.7,0.7}\polygon*(45,0)(110,0)(110,5)(45,5)}
      \put(35,30){\color[rgb]{0.7,0.7,0.7}\polygon*(0,0)(5,0)(5,5)(0,5)}
      \multiput(0,30)(5,0){30}{\polygon(0,0)(5,0)(5,5)(0,5)}
      \put(0,20){\color[rgb]{0.7,0.7,0.7}\polygon*(40,0)(105,0)(105,5)(40,5)}
      \put(110,20){\color[rgb]{0.7,0.7,0.7}\polygon*(0,0)(5,0)(5,5)(0,5)}
      \multiput(0,20)(5,0){30}{\polygon(0,0)(5,0)(5,5)(0,5)}
      \put(20,10){\color[rgb]{0.7,0.7,0.7}\polygon*(0,0)(5,0)(5,5)(0,5)}
      \put(35,10){\color[rgb]{0.7,0.7,0.7}\polygon*(0,0)(5,0)(5,5)(0,5)}
      \put(45,10){\color[rgb]{0.7,0.7,0.7}\polygon*(0,0)(5,0)(5,5)(0,5)}
      \put(0,10){\color[rgb]{0.7,0.7,0.7}\polygon*(55,0)(90,0)(90,5)(55,5)}
      \put(100,10){\color[rgb]{0.7,0.7,0.7}\polygon*(0,0)(5,0)(5,5)(0,5)}
      \put(115,10){\color[rgb]{0.7,0.7,0.7}\polygon*(0,0)(5,0)(5,5)(0,5)}
      \put(120,10){\color[rgb]{0.7,0.7,0.7}\polygon*(0,0)(5,0)(5,5)(0,5)}
      \put(130,10){\color[rgb]{0.7,0.7,0.7}\polygon*(0,0)(5,0)(5,5)(0,5)}
      \multiput(0,10)(5,0){30}{\polygon(0,0)(5,0)(5,5)(0,5)}
      \put(40,0){\color[rgb]{1,0,0}\line(0,1){46}}
      \put(110,0){\color[rgb]{1,0,0}\line(0,1){46}}
      \put(21.4,4.5){$\frac{7}{2}$}
      \put(36.4,4.5){$\frac{1}{2}$}
      \put(41.4,4.5){$\frac{1}{2}$}
      \put(51.4,4.5){$\frac{5}{2}$}
      \put(91.4,4.5){$\frac{7}{2}$}
      \put(96.4,4.5){$\frac{5}{2}$}
      \put(106.4,4.5){$\frac{1}{2}$}
      \put(116.4,4.5){$\frac{3}{2}$}
      \put(121.4,4.5){$\frac{5}{2}$}
      \put(131.4,4.5){$\frac{9}{2}$}
      \put(21,2){$\underbrace{\hspace{18\unitlength}}$}
      \put(41,2){$\underbrace{\hspace{13\unitlength}}$}
      \put(91,2){$\underbrace{\hspace{18\unitlength}}$}
      \put(111,2){$\underbrace{\hspace{23\unitlength}}$}
      \put(30,-4){$A^{-}$}
      \put(47.5,-4){$A_{0}^{-}$}
      \put(100,-4){$A_{0}^{+}$}
      \put(122.5,-4){$A^{+}$}
    \end{picture}
  \end{center}
  \caption{Choices of the numbers $k_{j}$, $j=1,\ldots,N$ characterizing some eigenstates. The gray squares represent the $k_{j}$'s chosen. The upper line corresponds to the choice for the stationary state (\ref{k0}). The second and third lines are the two possible choices for the gap. The last line corresponds to a generic eigenstate close to the stationary state, defined by sets $A_{0}^{-}=\{\frac{1}{2},\frac{5}{2}\}$, $A^{-}=\{\frac{1}{2},\frac{7}{2}\}$, $A_{0}^{+}=\{\frac{1}{2},\frac{5}{2},\frac{7}{2}\}$, $A^{+}=\{\frac{3}{2},\frac{5}{2},\frac{9}{2}\}$ of cardinals $m_{-}=|A_{0}^{-}|=|A^{-}|=2$ and $m_{+}=|A_{0}^{+}|=|A^{+}|=3$.}
  \label{fig choice k}
\end{figure}
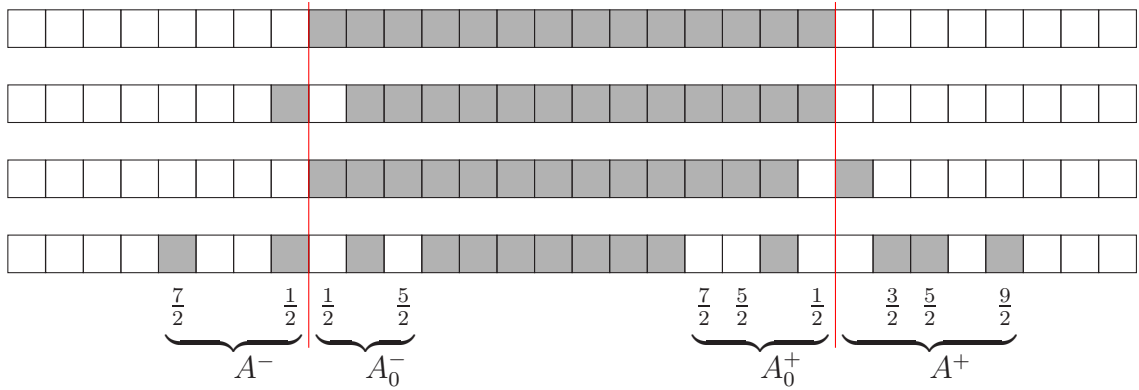

More generally, any excited state close to the stationary state can be defined by four finite sets of half integers describing which $k_{j}$'s are removed from both ends and where they are placed: the $k_{j}$'s removed are equal to $\pm(\rho L/2-a)$, $a\in A_{0}^{\pm}$, and they are placed at $\pm(\rho L/2+a)$, $a\in A^{\pm}$, with
\begin{equation}
\label{m+-}
m_{\pm}=|A_{0}^{\pm}|=|A^{\pm}|
\qquad
A_{0}^{\pm},A^{\pm}\subset\mathbb{N}+\frac{1}{2}\;.
\end{equation}
Then, for any function $f$, one can write
\begin{eqnarray}
\label{sum f(kj/L)}
\sum_{j=1}^{N}f\Big(\frac{k_{j}}{L}\Big)
=\sum_{j=1}^{N}f\Big(\frac{k_{j}^{(0)}}{L}\Big)
+\sum_{a\in A^{+}}f\Big(\frac{\rho}{2}+\frac{a}{L}\Big)
+\sum_{a\in A^{-}}f\Big(-\frac{\rho}{2}-\frac{a}{L}\Big)\\
\hspace{47mm}
-\sum_{a\in A_{0}^{+}}f\Big(\frac{\rho}{2}-\frac{a}{L}\Big)
-\sum_{a\in A_{0}^{-}}f\Big(-\frac{\rho}{2}+\frac{a}{L}\Big)\;.\nonumber
\end{eqnarray}
In particular, taking $f$ equal to the identity function and using (\ref{m+-}), the total momentum (\ref{P[k]}) rewrites
\begin{equation}
\label{P[A]}
P=\sum_{a\in A_{0}^{+}}a+\sum_{a\in A^{+}}a-\sum_{a\in A_{0}^{-}}a-\sum_{a\in A^{-}}a\;.
\end{equation}
We use again (\ref{sum f(kj/L)}) in section \ref{section asymptotic expansions} to extract the large $L$ asymptotics of $b$ and $E$ for the first excited states. A crucial point will be that both ends of the summation over $j$ in the right hand side of (\ref{sum f(kj/L)}) applied to (\ref{E[psi]}), (\ref{b[psi]}) correspond to branch points of the function $\psi$.
\end{subsection}

\begin{subsection}{Reformulation in terms of partitions of integers}
We define an index $Q$ equal to
\begin{equation}
\label{Q[A]}
Q=Q_{+}+Q_{-}
\quad
\text{with}
\quad
Q_{\pm}=\sum_{a\in A_{0}^{\pm}}a+\sum_{a\in A^{\pm}}a\;.
\end{equation}
The index $Q$ is useful in order to generate the first eigenvalues by increasing real part, see table \ref{table first eta c} and figure \ref{fig first eta c}.

We show in this section that the number $\Omega(Q)$ of eigenstates with a given value of $Q$ is equal to
\begin{equation}
\label{Omega(Q)}
\Omega(Q)=\sum_{q=0}^{Q}p(q)p(Q-q)\underset{Q\to\infty}{\simeq}\frac{3^{1/4}}{12Q^{5/4}}\,\rme^{2\pi\sqrt{Q/3}}\;,
\end{equation}
where $p(n)$ is the number of ways to partition $n$ as an unordered sum of positive integers. The numbers $\Omega(Q)$, $Q\in\mathbb{N}$ form the sequence A000712 of \cite{OEIS}. The large $Q$ behaviour follows from Hardy-Ramanujan asymptotics of $p(n)$ \cite{HR1918.1}.

In order to prove (\ref{Omega(Q)}), we construct a bijection between unordered partitions of $q$, and the ensemble of all possible choices for sets of positive half-integers $A_{0}$ and $A$ with same cardinal and verifying $q=\sum_{a\in A_{0}}a+\sum_{a\in A}a$. Calling $m$ the cardinal of $A_{0}$ and $A$, and respectively $a_{j}^{(0)}$ and $a_{j}$ the elements of $A_{0}$ and $A$ ordered from largest to smallest, one has
\begin{equation}
q=-m^{2}+\sum_{j=1}^{m}(a_{j}^{(0)}+j-\tfrac{1}{2})+\sum_{j=1}^{m}(a_{j}+j-\tfrac{1}{2})\;.
\end{equation}
The numbers $a_{j}^{(0)}+j-\frac{1}{2}$ and $a_{j}+j-\frac{1}{2}$ are integers larger or equal to $m$, non-increasing as $j$ increases. We define two partitions $\ell$ and $\overline{\ell}$ by $\ell_{j}=a_{j}^{(0)}+j-\frac{1}{2}$ and $\overline{\ell}_{j}=a_{j}+j-\frac{1}{2}$ for $1\leq j\leq m$ and $\ell_{j},\overline{\ell}_{j}\leq m$ for $j>m$. Adding the requirement that $\overline{\ell}$ is the conjugate partition of $\ell$ (the Ferrers diagram of $\overline{\ell}$ is obtained from the one of $\ell$ by exchanging the rows and the columns), this defines a unique pair $(\ell,\overline{\ell})$ of partitions, see figure \ref{fig  Ferrers diagram}. The number $q$ is then partitioned as $q=\ell_{1}+\ldots+\ell_{n}=\overline{\ell}_{1}+\ldots+\overline{\ell}_{\overline{n}}$ with $n=\overline{\ell}_{1}$ and $\overline{n}=\ell_{1}$. The cardinal $m$ of $A_{0}$ and $A$ corresponds to the size of the Durfee square of the partition (largest square that can fit in the Ferrers diagram).
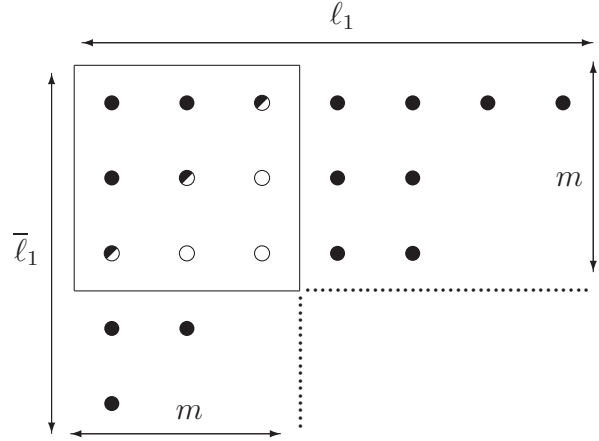
\begin{figure}
  \begin{center}
    \begin{tabular}{c}
    \begin{picture}(70,60)(-12,0)
      \multiput(5,45)(10,0){7}{\circle*{2}}
      \multiput(5,35)(10,0){5}{\circle*{2}}
      \multiput(5,25)(10,0){5}{\circle*{2}}
      \multiput(5,15)(10,0){2}{\circle*{2}}
      \multiput(5,05)(10,0){1}{\circle*{2}}
      \put(0,0){\color{white}\polygon*(1,20.9)(29,21)(29,48.9)}
      \multiput(25,45)(10,0){1}{\circle{2}}
      \multiput(15,35)(10,0){2}{\circle{2}}
      \multiput(5,25)(10,0){3}{\circle{2}}
      \put(0,50){\line(1,0){30}}
      \put(0,20){\line(1,0){30}}
      \put(0,50){\line(0,-1){30}}
      \put(30,50){\line(0,-1){30}}
      \put(69,50.5){\vector(0,-1){28}}
      \put(69,22.5){\vector(0,1){28}}
      \put(-3,49){\vector(0,-1){48}}
      \put(-3,01){\vector(0,1){48}}
      \put(-0.5,1){\vector(1,0){28}}
      \put(27.5,1){\vector(-1,0){28}}
      \put(01,53){\vector(1,0){68}}
      \put(69,53){\vector(-1,0){68}}
      \put(64,34){$m$}
      \put(13.5,3){$m$}
      \put(34,55.5){$\ell_{1}$}
      \put(-8,24){$\overline{\ell}_{1}$}
      \multiput(30.5,19)(1,0){38}{$\cdot$}
      \multiput(29.5,1)(0,1){18}{$\cdot$}
    \end{picture}
    \end{tabular}
  \end{center}
  \caption{Ferrers diagram associated to a partition $\ell$ of the integer $q=20$. Two sets of positive half-integers $A_{0}=\{a_{1}^{(0)},\ldots,a_{m}^{(0)}\}$ and $A=\{a_{1},\ldots,a_{m}\}$ are constructed from it, with $a_{j}^{(0)}$ equal to the number of filled dots in the $j$-th row, and $a_{j}$ equal to the number of filled dots in the $j$-th column. Half-filled dots count for $\frac{1}{2}$. The $m\times m$ Durfee square of the Ferrers diagram is represented in the picture.}
  \label{fig  Ferrers diagram}
\end{figure}

In the rest of the article, we write everything in terms of the sets $A_{0}^{\pm}$, $A^{\pm}$. The translation to integer partitions $\ell^{\pm}$ and their conjugates $\overline{\ell^{\pm}}$ can be done with the following formulas, valid for an arbitrary function $f$:
\begin{eqnarray}
\sum_{a\in A_{0}^{+}}f(a)=\sum_{j=1}^{m_{+}}f(\ell_{j}^{+}-j+\tfrac{1}{2})\qquad
\sum_{a\in A^{+}}f(a)=\sum_{j=1}^{m_{+}}f(\overline{\ell^{+}_{j}}-j+\tfrac{1}{2})\\
\sum_{a\in A_{0}^{-}}f(a)=\sum_{j=1}^{m_{-}}f(\ell_{j}^{-}-j+\tfrac{1}{2})\qquad
\sum_{a\in A^{-}}f(a)=\sum_{j=1}^{m_{-}}f(\overline{\ell_{j}^{-}}-j+\tfrac{1}{2})\;.\nonumber
\end{eqnarray}
\end{subsection}
\end{section}

%%%%%%%%%%%%%%%%%
%%             %%
%%  Section 3  %%
%%             %%
%%%%%%%%%%%%%%%%%
\begin{section}{Asymptotic expansions}
\label{section asymptotic expansions}
In this section, we calculate the large $L$ asymptotic expansions of equations (\ref{b[psi]}) and (\ref{E[psi]}) for $b$ and $E$ for the first excited states. More generally, we also extract the asymptotic expansion of the eigenvalues of the transfer matrix related to TASEP from quantum integrability.

\begin{subsection}{Asymptotic expansion of the parameter \texorpdfstring{$b$}{b}}
\label{subsection expansion b}
We show that for the first excited states, the parameter $b$ defined in (\ref{b[y]}) is asymptotically equal to $b_{0}$ for large $L$, with
\begin{equation}
\label{b0}
b_{0}=\rho\log\rho+(1-\rho)\log(1-\rho)<0\;.
\end{equation}
In order to do this, we write
\begin{equation}
\label{b[c]}
b=b_{0}+\frac{2\pi c}{L}\;.
\end{equation}
We assume in the following that $\Re\,c>0$, which is sufficient to show that $b\simeq b_{0}$ and to derive the asymptotic expressions (\ref{eq eta' c}), (\ref{E[c]}). We argue in section \ref{subsection approximation eta} that one has in fact the stronger constraint $-\frac{\pi}{6}<\arg c<\frac{\pi}{6}$.

The sum over $j$ in the equation (\ref{b[psi]}) for $b$ can be expressed as in (\ref{sum f(kj/L)}). Using (\ref{k0}), one has
\begin{eqnarray}
\label{b[psi,A,L]}
b_{0}+\frac{2\pi c}{L}
=\frac{1}{L}\sum_{j=1}^{N}\psi\Big(\rme^{-\rmi\pi\rho-b_{0}+\frac{2\rmi\pi(j-1/2+\rmi c)}{L}}\Big)\\
\hspace{21mm}
+\frac{1}{L}\sum_{a\in A^{+}}\psi\Big(\rme^{\rmi\pi\rho-b_{0}-\frac{2\pi(c-\rmi a)}{L}}\Big)
+\frac{1}{L}\sum_{a\in A^{-}}\psi\Big(\rme^{-\rmi\pi\rho-b_{0}-\frac{2\pi(c+\rmi a)}{L}}\Big)\nonumber\\
\hspace{21mm}
-\frac{1}{L}\sum_{a\in A_{0}^{+}}\psi\Big(\rme^{\rmi\pi\rho-b_{0}-\frac{2\pi(c+\rmi a)}{L}}\Big)
-\frac{1}{L}\sum_{a\in A_{0}^{-}}\psi\Big(\rme^{-\rmi\pi\rho-b_{0}-\frac{2\pi(c-\rmi a)}{L}}\Big)\;.\nonumber
\end{eqnarray}
The asymptotic expansion of the four sums over $a$ in (\ref{b[psi,A,L]}) can be computed from (\ref{psi(exp(...-epsilon))}). Up to order $L^{-5/2}$, using (\ref{m+-}) to cancel a few terms, we find that their sum is equal to
\begin{eqnarray}
-\frac{2\rmi\sqrt{\pi}}{\sqrt{\rho(1-\rho)}L^{3/2}}\Big(\sum_{a\in A_{0}^{+}}\sqrt{c+\rmi a}+\sum_{a\in A^{-}}\sqrt{c+\rmi a}\nonumber\\
\hspace{50mm}
-\sum_{a\in A_{0}^{-}}\sqrt{c-\rmi a}-\sum_{a\in A^{+}}\sqrt{c-\rmi a}\Big)\nonumber\\
-\frac{2\rmi\pi(1-2\rho)}{3\rho(1-\rho)L^{2}}\Big(\sum_{a\in A_{0}^{+}}a+\sum_{a\in A^{+}}a-\sum_{a\in A_{0}^{-}}a-\sum_{a\in A^{-}}a\Big)\\
+\frac{2\rmi\pi^{3/2}(1-\rho+\rho^{2})}{9(\rho(1-\rho))^{3/2}L^{5/2}}\Big(\sum_{a\in A_{0}^{+}}(c+\rmi a)^{3/2}+\sum_{a\in A^{-}}(c+\rmi a)^{3/2}\nonumber\\
\hspace{50mm}
-\sum_{a\in A_{0}^{-}}(c-\rmi a)^{3/2}-\sum_{a\in A^{+}}(c-\rmi a)^{3/2}\Big)\;.\nonumber
\end{eqnarray}
For the sum over $j$ in (\ref{b[psi,A,L]}), one has to use a formulation of the Euler-Maclaurin asymptotic expansion (see \textit{e.g.} \cite{H1949.1}) with square root singularities at both ends of the summation range (\ref{S[zeta]}), with $f(x)=\psi(\rme^{-\rmi\pi\rho-b_{0}+2\rmi\pi x})$ and $d=-\frac{1}{2}+\rmi c$. Indeed, from (\ref{psi(exp(...-epsilon))}) and using $\sqrt{-\rmi x}=\sqrt{-\rmi}\sqrt{x}$ for $-\pi/2<\arg x<\pi$, the expansion for small $x$ of $f(x)$ is of the form (\ref{f[f,fbar]}) with
\begin{eqnarray}
f_{0}=\log\Big(\frac{\rho}{1-\rho}\Big)+\rmi\pi\nonumber\\
f_{1}=-\frac{(1+\rmi)\sqrt{2\pi}}{\sqrt{\rho(1-\rho)}}\nonumber\\
f_{2}=-\frac{2\rmi\pi(1-2\rho)}{3\rho(1-\rho)}\\
f_{3}=\frac{\sqrt{2}\pi^{3/2}(1-\rmi)(1-\rho+\rho^{2})}{9(\rho(1-\rho))^{3/2}}\;\ldots\nonumber
%f_{4}=\frac{4\pi^{2}(1+\rho)(2-\rho)(1-2\rho)}{135\rho^{2}(1-\rho)^{2}}\;.\nonumber
\end{eqnarray}
For the expansion near $x=\rho$, $-\pi<\arg(\rho-x)<\pi/2$, we find that the coefficients $\bar{f}_{k}$ of (\ref{f[f,fbar]}) are the complex conjugates of the $f_{k}$. Under the assumption $\Re\,c>0$, the conditions below (\ref{f[f,fbar]}) are verified and (\ref{S[zeta]}) implies the asymptotic expansion
\begin{eqnarray}
\label{Euler-Maclaurin b}
\frac{1}{L}\sum_{j=1}^{N}\psi\Big(\rme^{-\rmi\pi\rho-b_{0}+\frac{2\rmi\pi(j-1/2+\rmi c)}{L}}\Big)
\simeq\Big(\int_{-\rho/2}^{\rho/2}\rmd u\,\psi(\rme^{2\rmi\pi u-b_{0}})\Big)\\
\hspace{20mm}
+\sum_{k=0}^{\infty}\frac{f_{k}}{L^{1+k/2}}\,\zeta(-\tfrac{k}{2},\,\tfrac{1}{2}+\rmi c)
+\sum_{k=0}^{\infty}\frac{\bar{f}_{k}}{L^{1+k/2}}\,\zeta(-\tfrac{k}{2},\,\tfrac{1}{2}-\rmi c)\;,\nonumber
\end{eqnarray}
where $\zeta$ is Hurwitz zeta function (\ref{zeta def}). For $k$ even, the $\zeta$ can be replaced by Bernoulli polynomials defined in (\ref{Br(z)}). Using $B_{1}(x)=x-1/2$, the term $k=0$ gives $2\pi c/L$, which cancels part of the left hand side of (\ref{b[psi,A,L]}). The terms $\zeta(-k/2,1/2\pm\rmi c)$ with $k=2$ cancel because of the symmetry (\ref{Br(1-z)}).

The integral in (\ref{Euler-Maclaurin b}) is equal to $b_{0}$. Indeed, using (\ref{psi(exp(...))}), (\ref{psi}) and (\ref{psi'}), the change of variables $z=\psi(\rme^{2\rmi\pi u-b_{0}})$ leads to
\begin{equation}
\int_{-\rho/2}^{\rho/2}\rmd u\,\psi(\rme^{2\rmi\pi u-b_{0}})
=\int_{\log(\frac{\rho}{1-\rho})+\rmi\pi}^{\log(\frac{\rho}{1-\rho})-\rmi\pi}\frac{\rmd z}{2\rmi\pi}\,\Big[\frac{z\,\rme^{z}}{\rme^{z}-1}-\rho z\Big]\;.
\end{equation}
The integral can be performed explicitly in terms of the dilogarithm function $\Li_{2}$, using $\partial_{z}[z\log(1-\rme^{z})+\Li_{2}(\rme^{z})]=z\,\rme^{z}/(\rme^{z}-1)$. We note that $\rme^{z}$ has the same value at both ends of the integration range. Numerically, we observe that the contour for $z$ does not cross the line $[1,\infty[$ (the contour for $1-z$ is equal to the boundary of the domain $\mathcal{B}_{-}$ in figure \ref{fig B}), which corresponds to the cut of $\Li_{2}$ and to the cut of $y\mapsto\log(1-y)$. It implies that the dilogarithm cancels, and that we can take the same branch of the logarithm in $\log(1-\rme^{z})$ for all $z$. One finally finds that the integral is equal to the definition (\ref{b0}) of $b_{0}$. Thus, the integral in (\ref{Euler-Maclaurin b}) cancels a term in the left hand side of (\ref{b[psi,A,L]}).

Finally, gathering everything and using $\partial_{x}\zeta(s,x)=-s\zeta(s+1,x)$, we find that the expansion up to order $L^{-5/2}$ of (\ref{b[psi,A,L]}) gives
\begin{equation}
\label{eq eta' c subleading terms}
\eta'(c)=\frac{4\sqrt{2}\,\rmi\pi^{5/2}(1-2\rho)}{3\sqrt{\rho(1-\rho)}}\,\frac{P}{\sqrt{L}}+\frac{\pi(1-\rho+\rho^{2})}{6\rho(1-\rho)}\,\frac{\eta(c)}{L}+\O{L^{-3/2}}\;,
\end{equation}
with
\begin{eqnarray}
\label{eta[A,zeta]}
\begin{picture}(0,0)(0,-8.5)
  \put(0,0){\line(1,0){105.5}}
  \put(0,-40){\line(1,0){105.5}}
  \put(0,0){\line(0,-1){40}}
  \put(105.5,0){\line(0,-1){40}}
\end{picture}
\hspace{2mm}
\eta(c)=-\frac{8\pi^{2}}{3}\Big((1-\rmi)\zeta(-\tfrac{3}{2},\tfrac{1}{2}+\rmi c)+(1+\rmi)\zeta(-\tfrac{3}{2},\tfrac{1}{2}-\rmi c)\Big)\\
\hspace{14.7mm}
-\frac{8\sqrt{2}\,\rmi\pi^{2}}{3}\Big(\sum_{a\in A_{0}^{+}}(c+\rmi a)^{3/2}+\sum_{a\in A^{-}}(c+\rmi a)^{3/2}\nonumber\\
\hspace{40mm}
-\sum_{a\in A_{0}^{-}}(c-\rmi a)^{3/2}-\sum_{a\in A^{+}}(c-\rmi a)^{3/2}\Big)\;\nonumber
\end{eqnarray}
and $P$ defined by (\ref{P[A]}). Thus, at leading order in $L$, the equation for $c$ is simply
\begin{equation}
\label{eq eta' c}
\boxed{\eta'(c)=0}\;.
\end{equation}
We conjecture that (\ref{eq eta' c}) has a unique solution in the domain $\Re\,c>0$. This seems confirmed by numerical checks for the first eigenstates.

The subleading terms in (\ref{eq eta' c subleading terms}) are necessary in the next subsection to extract from (\ref{E[psi]}) the leading behaviour (\ref{E[c]}) of the eigenvalue $E$.
\end{subsection}

\begin{subsection}{Asymptotic expansion of the eigenvalues of the Markov matrix}
\label{subsection expansion E}
We write again $b$ as (\ref{b[c]}). The sum over $j$ in the formula (\ref{E[psi]}) for $E$ can also be expressed as in (\ref{sum f(kj/L)}):
\begin{eqnarray}
\label{E[psi,A,L]}
E=\sum_{j=1}^{N}\Big(\exp\Big[-\psi\Big(\rme^{-\rmi\pi\rho-b_{0}+\frac{2\rmi\pi(j-1/2+\rmi c)}{L}}\Big)\Big]-1\Big)^{-1}\nonumber\\
\hspace{8mm}
+\sum_{a\in A^{+}}\Big(\exp\Big[-\psi\Big(\rme^{\rmi\pi\rho-b_{0}-\frac{2\pi(c-\rmi a)}{L}}\Big)\Big]-1\Big)^{-1}\nonumber\\
\hspace{8mm}
+\sum_{a\in A^{-}}\Big(\exp\Big[-\psi\Big(\rme^{-\rmi\pi\rho-b_{0}-\frac{2\pi(c+\rmi a)}{L}}\Big)\Big]-1\Big)^{-1}\\
\hspace{8mm}
-\sum_{a\in A_{0}^{+}}\Big(\exp\Big[-\psi\Big(\rme^{\rmi\pi\rho-b_{0}-\frac{2\pi(c+\rmi a)}{L}}\Big)\Big]-1\Big)^{-1}\nonumber\\
\hspace{8mm}
-\sum_{a\in A_{0}^{-}}\Big(\exp\Big[-\psi\Big(\rme^{-\rmi\pi\rho-b_{0}-\frac{2\pi(c-\rmi a)}{L}}\Big)\Big]-1\Big)^{-1}\;.\nonumber
\end{eqnarray}
The asymptotic expansion of the four sums over $a$ in (\ref{E[psi,A,L]}) can be computed from (\ref{psi(exp(...-epsilon))}). Up to order $L^{-3/2}$ one finds, using (\ref{m+-}) to cancel a few terms, that their sum is equal to
\begin{eqnarray}
\frac{2\rmi\sqrt{\pi}\sqrt{\rho(1-\rho)}}{\sqrt{L}}\Big(\sum_{a\in A_{0}^{+}}\sqrt{c+\rmi a}+\sum_{a\in A^{-}}\sqrt{c+\rmi a}\nonumber\\
\hspace{50mm}
-\sum_{a\in A_{0}^{-}}\sqrt{c-\rmi a}-\sum_{a\in A^{+}}\sqrt{c-\rmi a}\Big)\nonumber\\
-\frac{4\rmi\pi(1-2\rho)}{3L}\Big(\sum_{a\in A_{0}^{+}}a+\sum_{a\in A^{+}}a-\sum_{a\in A_{0}^{-}}a-\sum_{a\in A^{-}}a\Big)\\
-\frac{2\rmi\pi^{3/2}(1-13\rho+13\rho^{2})}{9\sqrt{\rho(1-\rho)}L^{3/2}}\Big(\sum_{a\in A_{0}^{+}}(c+\rmi a)^{3/2}+\sum_{a\in A^{-}}(c+\rmi a)^{3/2}\nonumber\\
\hspace{60mm}
-\sum_{a\in A_{0}^{-}}(c-\rmi a)^{3/2}-\sum_{a\in A^{+}}(c-\rmi a)^{3/2}\Big)\;.\nonumber
\end{eqnarray}
For the term with a sum over $j$ in (\ref{E[psi,A,L]}), one can use again Euler-Maclaurin formula with square root singularities at both ends (\ref{S[zeta]}) for $f(x)=(\exp(-\psi(\rme^{-\rmi\pi\rho-b_{0}+2\rmi\pi x}))-1)^{-1}$ and $d=-\frac{1}{2}+\rmi c$. Indeed, from (\ref{psi(exp(...-epsilon))}), the expansion for small $x$, $-\pi/2<\arg x<\pi$, of $f(x)$ is of the form (\ref{f[f,fbar]}) with
\begin{eqnarray}
f_{0}=-\rho\nonumber\\
f_{1}=(1+\rmi)\sqrt{2\pi}\sqrt{\rho(1-\rho)}\nonumber\\
f_{2}=-\frac{4\rmi\pi(1-2\rho)}{3}\\
f_{3}=-\frac{\sqrt{2}\pi^{3/2}(1-\rmi)(1-13\rho+13\rho^{2})}{9\sqrt{\rho(1-\rho)}}\;\ldots\nonumber
%f_{4}=-\frac{8\pi^{2}(1-2\rho)(1+23\rho-23\rho^{2})}{135\rho(1-\rho)}\;.\nonumber
\end{eqnarray}
For the expansion near $x=\rho$, $-\pi<\arg(\rho-x)<\pi/2$, we find that the coefficients $\bar{f}_{k}$ are again the complex conjugates of the $f_{k}$. From (\ref{S[zeta]}), one has the asymptotic expansion
\begin{eqnarray}
\label{Euler-Maclaurin E}
\sum_{j=1}^{N}\Big(\exp\Big[-\psi\Big(\rme^{-\rmi\pi\rho-b_{0}+\frac{2\rmi\pi(j-1/2+\rmi c)}{L}}\Big)\Big]-1\Big)^{-1}\\
\hspace{10mm}
\simeq L\Big(\int_{-\rho/2}^{\rho/2}\rmd u\,\Big(\exp[-\psi(\rme^{2\rmi\pi u-b_{0}})]-1\Big)^{-1}\Big)\nonumber\\
\hspace{20mm}
+\sum_{k=0}^{\infty}\frac{f_{k}}{L^{k/2}}\,\zeta(-\tfrac{k}{2},\,\tfrac{1}{2}+\rmi c)
+\sum_{k=0}^{\infty}\frac{\bar{f}_{k}}{L^{k/2}}\,\zeta(-\tfrac{k}{2},\,\tfrac{1}{2}-\rmi c)\;.\nonumber
\end{eqnarray}
For $k=0$ and $k=2$, the terms $\zeta(-k/2,1/2\pm\rmi c)$ cancel from (\ref{Br(z)}), (\ref{Br(1-z)}).

The integral in (\ref{Euler-Maclaurin E}) is equal to $0$. Indeed, using again (\ref{psi(exp(...))}), (\ref{psi}) and (\ref{psi'}), the change of variables $z=\psi(\rme^{2\rmi\pi u-b_{0}})$ leads to
\begin{equation}
\int_{-\rho/2}^{\rho/2}\frac{\rmd u}{\exp[-\psi(\rme^{2\rmi\pi u-b_{0}})]-1}
=\int_{\log(\frac{\rho}{1-\rho})+\rmi\pi}^{\log(\frac{\rho}{1-\rho})-\rmi\pi}\!\frac{\rmd z}{2\rmi\pi}\,\Big[\frac{\rho\,\rme^{z}}{\rme^{z}-1}-\frac{\rme^{2z}}{(\rme^{z}-1)^{2}}\Big]\;.
\end{equation}
Unlike in the corresponding calculation for $b$, the integrand is now a single-valued function of $\rme^{z}$. The change of variables $y=\rme^{z}$ gives the integration of $\rho/(y-1)-y/(y-1)^{2}$ over the contour $\mathcal{C}=\{\exp(\psi(\rme^{2\rmi\pi u-b_{0}})),u\in[-\rho/2,\rho/2]\}$. Numerically, we observe that this contour does not enclose $1$ (in figure \ref{fig B}, the contour for $1-y$ is the boundary of the domain $\mathcal{B}_{-}$). The integral thus vanishes, as expected since $E=0$ for the stationary state.

Finally, gathering everything and using (\ref{eq eta' c subleading terms}) to cancel the terms of order $1/\sqrt{L}$, one finds
\begin{equation}
\label{E[c]}
\boxed{E\simeq-\frac{2\rmi\pi(1-2\rho)}{L}\,P-\frac{\sqrt{\rho(1-\rho)}}{\sqrt{2\pi}L^{3/2}}\,\eta(c)}\;,
\end{equation}
where the parameter $c$ is the solution of (\ref{eq eta' c}), and $P$ and $\eta$ are defined respectively in (\ref{P[A]}) and (\ref{eta[A,zeta]}). We observe that the term of order $L^{-3/2}$ of the eigenvalue is given by the value of the function $\eta$ at its saddle point, that we conjectured to be unique in the domain $\Re\,c>0$.

The eigenvalues are very degenerate, as already observed in \cite{GM2004.2,GM2005.2}. Indeed, from the expression (\ref{eta[A,zeta]}) of $\eta$, exchanging elements between $A_{0}^{+}$ and $A^{-}$, and between $A_{0}^{-}$ and $A^{+}$ (with the constraints $|A_{0}^{+}|=|A^{+}|$ and $|A_{0}^{-}|=|A^{-}|$) does not change the function $\eta$. At half-filling $\rho=1/2$, the eigenvalue is preserved by such an exchange, but not necessarily the total momentum $P$.
\end{subsection}

\begin{subsection}{Asymptotic expansion of the eigenvalues of the transfer matrix}
\label{subsection expansion E(lambda)}
The eigenvalues $\mathcal{E}(\lambda)$ of the transfer matrix $t(\lambda)$, related to the Markov matrix by $M=t'(0)t^{-1}(0)$, are written in terms of the Bethe roots as (see \cite{GM2006.2} equation (21) for TASEP, with $z_{j}=1-y_{j}$)
\begin{equation}
\mathcal{E}(\lambda)=\prod_{j=1}^{N}\frac{1}{1-(1-\lambda)^{-1}y_{j}}+\lambda^{L}\,\prod_{j=1}^{N}\frac{1}{1-(1-\lambda)y_{j}^{-1}}\;.
\end{equation}
Writing $y_{j}$ as (\ref{y[k,b]}) and using (\ref{b[psi]}), one has
\begin{eqnarray}
\label{E[lambda,psi] 1}
\mathcal{E}(\lambda)=\Big(1+\frac{(-1)^{N}\lambda^{L}\rme^{bL}}{(1-\lambda)^{N}}\Big)\exp\Bigg(-\sum_{j=1}^{N}\log\Big(1-\frac{\rme^{\psi\big(\rme^{\frac{2\rmi\pi k_{j}}{L}-b}\big)}}{1-\lambda}\Big)\Bigg)\\
\label{E[lambda,psi] 2}
\hspace{10mm}
=\lambda^{L}\Big(1+\frac{(1-\lambda)^{N}}{(-1)^{N}\lambda^{L}\rme^{bL}}\Big)\exp\Bigg(-\sum_{j=1}^{N}\log\Big(1-\frac{1-\lambda}{\rme^{\psi\big(\rme^{\frac{2\rmi\pi k_{j}}{L}-b}\big)}}\Big)\Bigg)\;.
\end{eqnarray}
In order to extract the large $L$ limit with fixed $\lambda$ of $\mathcal{E}(\lambda)$ for the first eigenstates, one writes again $b$ as in (\ref{b[c]}). Several cases for $\lambda$ must be treated separately. First, the prefactors of the exponential in (\ref{E[lambda,psi] 1}) and (\ref{E[lambda,psi] 2}) depend for large $L$ on whether $|\lambda(1-\lambda)^{-\rho}\rme^{b_{0}}|$ is smaller or larger than $1$. Furthermore, in order to avoid the cut of the logarithm in the Riemann integral obtained from the sum over $j$ with $k_{j}=k_{j}^{(0)}$, we need to choose either expression (\ref{E[lambda,psi] 1}) or (\ref{E[lambda,psi] 2}) for $\mathcal{E}(\lambda)$.

The contour line for the variable $\lambda$ of $|g(1-\lambda)|=\rme^{-b_{0}}$ with $g$ defined in (\ref{g}) partitions the complex plane into $3$ non overlapping domains $\mathcal{B}_{-}$, $\mathcal{B}_{+}$ and $\overline{\mathcal{B}}$, see figure \ref{fig B}. The border between the domains is the curve $\{1-\exp(\psi(\rme^{2\rmi\pi u-b_{0}})),0\leq u\leq1\}$. The border of the domain $\mathcal{B}_{+}$ intersects the real axis at the points $\lambda_{0}=(1-\rho)^{-1}$ and $\lambda_{\pm}=1-\exp(\psi(\pm\rme^{-b_{0}}))$. They verify $\lambda_{-}<-1<\frac{1}{2}<\lambda_{+}<1<\lambda_{0}$ for $0<\rho<1$.
\begin{figure}
  \begin{center}
    \begin{tabular}{c}
    \begin{picture}(0,0)
      \put(62.5,41.5){$\mathcal{B}_{-}$}
      \put(25,59){$\mathcal{B}_{+}$}
      \put(7,72){$\overline{\mathcal{B}}$}
    \end{picture}
    \includegraphics[width=75mm]{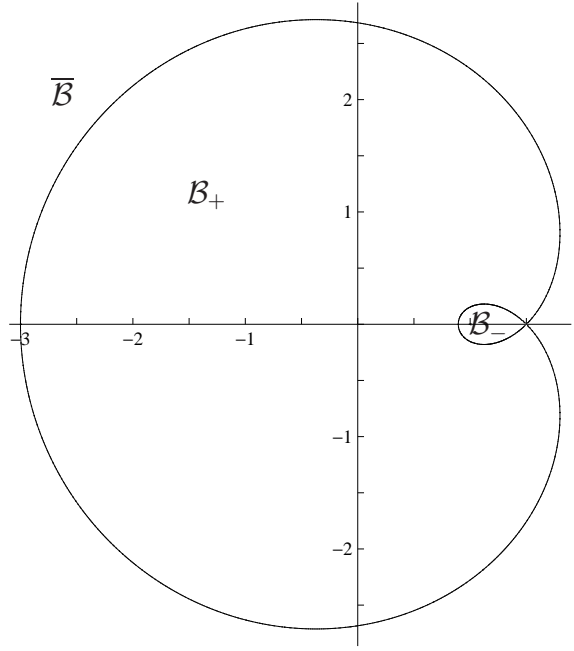}
    \end{tabular}
  \end{center}
  \caption{Curve of the points $\lambda$ in the complex plane verifying $|g(1-\lambda)|=\rme^{-b_{0}}$ with $g$ defined in (\ref{g}), $b_{0}$ defined in (\ref{b0}) and $\rho=1/3$. Equivalently, curve of $1-\exp(\psi(\rme^{2\rmi\pi u-b_{0}}))$, $0\leq u\leq1$ with $\psi$ defined in (\ref{psi}). This curve defines a partition of the complex plane into $3$ domains $\mathcal{B}_{-}$, $\mathcal{B}_{+}$ and $\overline{\mathcal{B}}$.}
  \label{fig B}
\end{figure}

The border of the domain $\mathcal{B}_{-}$ is $\{1-\exp(\psi(\rme^{2\rmi\pi u-b_{0}})),-\frac{\rho}{2}\leq u\leq\frac{\rho}{2}\}$. Since $1$ always belongs to $\mathcal{B}_{-}$ and $\mathcal{B}_{-}$ is a convex set, one has $\mathcal{B}_{-}=\{1-r\exp(\psi(\rme^{2\rmi\pi u-b_{0}})),-\frac{\rho}{2}\leq u\leq\frac{\rho}{2},0\leq r<1\}$: the condition $\lambda\in\mathcal{B}_{-}$ is equivalent to $|(1-\lambda)\exp(-\psi(\rme^{2\rmi\pi u-b_{0}}))|<1$ for $-\frac{\rho}{2}\leq u\leq\frac{\rho}{2}$, so that in the Riemann integral coming from (\ref{E[lambda,psi] 2}), the cut of the logarithm is never crossed. Similarly, one has $\mathcal{B}_{+}\cup\overline{\mathcal{B}}=\{1-r\exp(\psi(\rme^{2\rmi\pi u-b_{0}})),-\frac{\rho}{2}\leq u\leq\frac{\rho}{2},1<r\}$, so that for $\lambda\in\mathcal{B}_{+}\cup\overline{\mathcal{B}}$, it is convenient to use (\ref{E[lambda,psi] 1}) to avoid the cut of the logarithm.

We also observe that the condition $|\lambda(1-\lambda)^{-\rho}\rme^{b_{0}}|<1$ is equivalent to $\lambda\in\mathcal{B}_{+}$. Then, up to terms exponentially small when $L\to\infty$, one has
\begin{eqnarray}
& \mathcal{E}(\lambda)\simeq
\lambda^{L}\exp\Big(-\sum_{j=1}^{N}\log\Big(1-\frac{1-\lambda}{\rme^{\psi\big(\rme^{\frac{2\rmi\pi k_{j}}{L}-b}\big)}}\Big)\Big)
& \quad\lambda\in\mathcal{B}_{-}\\
& \mathcal{E}(\lambda)\simeq
\exp\Big(-\sum_{j=1}^{N}\log\Big(1-\frac{\rme^{\psi\big(\rme^{\frac{2\rmi\pi k_{j}}{L}-b}\big)}}{1-\lambda}\Big)\Big)
& \quad\lambda\in\mathcal{B}_{+}\\
& \mathcal{E}(\lambda)\simeq
\frac{(-1)^{N}\lambda^{L}\rme^{bL}}{(1-\lambda)^{N}}\,\exp\Big(-\sum_{j=1}^{N}\log\Big(1-\frac{\rme^{\psi\big(\rme^{\frac{2\rmi\pi k_{j}}{L}-b}\big)}}{1-\lambda}\Big)\Big)
& \quad\lambda\in\overline{\mathcal{B}}\;.
\end{eqnarray}
One can then write the sums over $j$ as in (\ref{sum f(kj/L)}), and use again Euler-Maclaurin formula. At leading order in $L$, the integrals can be computed by the same change of variables as in section \ref{subsection expansion E}. We find
\begin{eqnarray}
& \int_{-\rho-2}^{\rho/2}\rmd u\,\log\Big(1-\frac{1-\lambda}{\rme^{\psi(\rme^{2\rmi\pi u-b_{0}})}}\Big)=\log\lambda
& \quad\lambda\in\mathcal{B}_{-}\\
& \int_{-\rho-2}^{\rho/2}\rmd u\,\log\Big(1-\frac{\rme^{\psi(\rme^{2\rmi\pi u-b_{0}})}}{1-\lambda}\Big)=0
& \quad\lambda\in\mathcal{B}_{+}\cup\overline{\mathcal{B}}\;.
\end{eqnarray}
After some calculations, one finally obtains the same expansion for all $\lambda\in\mathcal{B}=\mathcal{B}_{-}\cup\mathcal{B}_{+}$:
\begin{equation}
\label{E[lambda,c] 1}
\boxed{\log\mathcal{E}(\lambda)\simeq-\frac{2\rmi\pi(1-\lambda)}{(1-(1-\rho)\lambda)^{2}}\,\frac{P}{L}-\frac{\sqrt{\rho(1-\rho)}\lambda(1-\lambda)}{\sqrt{2\pi}(1-(1-\rho)\lambda)^{3}}\,\frac{\eta(c)}{L^{3/2}}}\;.
\end{equation}
For $\lambda=0\in\mathcal{B}_{+}$, we recover (\ref{E[c]}) from $E=\mathcal{E}'(0)/\mathcal{E}(0)$. One can also extract from (\ref{E[lambda,c] 1}) the eigenvalues of the generalized Hamiltonians $F_{k}$ defined in section \ref{subsection transfer matrix}.

The calculation for $\lambda\in\overline{\mathcal{B}}$ is exactly the same as for $\lambda\in\mathcal{B}_{+}$, except for the extra factor which is exponentially large with $L$. One finds
\begin{equation}
\label{E[lambda,c] 2}
\fl\hspace{4mm}
\boxed{\log\Big(\frac{(\lambda-1)^{N}\mathcal{E}(\lambda)}{\lambda^{L}\rme^{bL}}\Big)\simeq-\frac{2\rmi\pi(1-\lambda)}{(1-(1-\rho)\lambda)^{2}}\,\frac{P}{L}-\frac{\sqrt{\rho(1-\rho)}\lambda(1-\lambda)}{\sqrt{2\pi}(1-(1-\rho)\lambda)^{3}}\,\frac{\eta(c)}{L^{3/2}}}\;.
\end{equation}
\end{subsection}
\end{section}

%%%%%%%%%%%%%%%%%
%%             %%
%%  Section 4  %%
%%             %%
%%%%%%%%%%%%%%%%%
\begin{section}{The function \texorpdfstring{$\eta$}{eta}}
\label{section eta}
In this section, we discuss some properties of the function $\eta$, defined in (\ref{eta[A,zeta]}), and in terms of which are expressed the eigenvalues of the Markov matrix (\ref{E[c]}), the eigenvalues of the transfer matrix (\ref{E[lambda,c] 1}), (\ref{E[lambda,c] 2}) and the parameter $c$ (\ref{eq eta' c}) for the first excited states.

\begin{subsection}{Branch cuts}
Hurwitz zeta function $\zeta(-3/2,z)$ has a branch point at $z=0$. The branch cut is chosen equal to $\mathbb{R}^{-}$ as usual. Then, $\zeta(-3/2,z)$ is analytic in $\mathbb{C}\backslash\mathbb{R}^{-}$. The term of $\eta(c)$ with the $\zeta$ functions has then branch cuts $\pm\rmi[1/2,\infty[$, while the terms $(c\pm\rmi a)^{3/2}$ correspond to additional branch cuts $\mathbb{R}^{-}\mp\rmi a$. These last branch cuts can always be rotated by an angle $\pm\pi/2$ around the branch point, so that the new branch cuts are now included in the branch cut coming from $\zeta$: this rotation does not change the value of $\eta(c)$ when $\Re\,c>0$.

It follows that $\eta(c)$ is analytic in the domain $\Re\,c>0$. Contour lines of the function $\eta$ are plotted for one particular eigenstate in figure \ref{fig eta}.
\begin{figure}
  \begin{center}
    \begin{tabular}{cc}
      \begin{tabular}{c}
      \includegraphics[width=70mm]{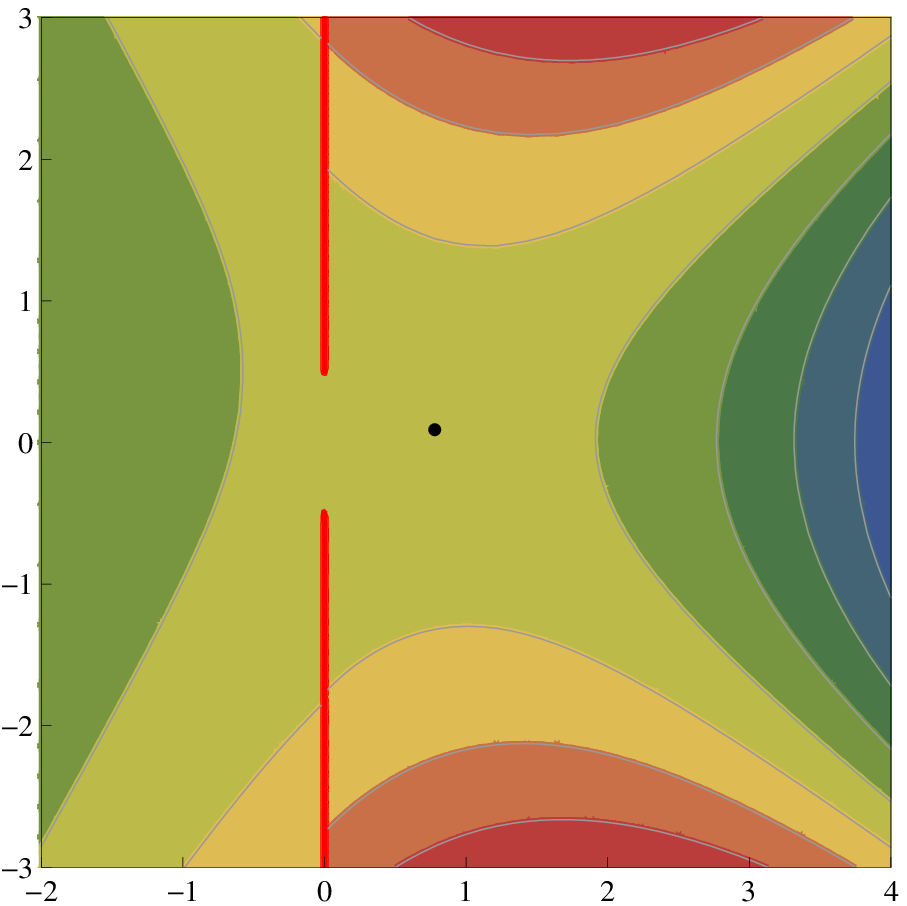}
      \end{tabular}
      &
      \begin{tabular}{c}
      \includegraphics[width=70mm]{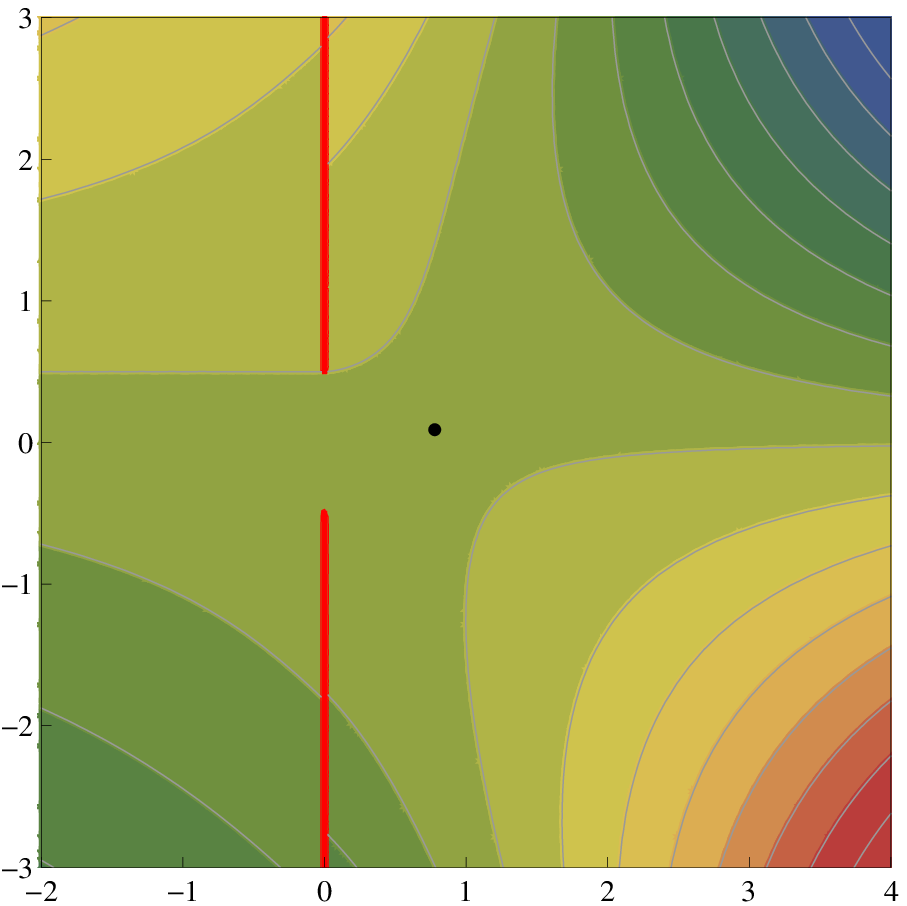}
      \end{tabular}
    \end{tabular}
  \end{center}
  \caption{Contour line of the real part (left) and the imaginary part (right) of the function $\eta$, defined in (\ref{eta[A,zeta]}), for $A_{0}^{+}=\{\frac{1}{2}\}$, $A^{+}=\{\frac{3}{2}\}$, $A_{0}^{-}=A^{-}=\{\}$. The red, straight lines are a possible choice for the branch cuts of $\eta$. The dot in the middle is the position of the saddle point $c\approx0.778+0.09\,\rmi$ of $\eta$, solution of $\eta'(c)=0$.}
  \label{fig eta}
\end{figure}
\end{subsection}

\begin{subsection}{Relation with polylogarithms}
The function $\eta$ can be expressed in terms of a polylogarithm function instead of Hurwitz zeta function, using Jonqui\`ere's identity
\begin{equation}
\label{zeta[Li]}
\zeta(1-s,q)
=\frac{\Gamma(s)}{(2\pi)^{s}}
\Big(
\rme^{-\frac{\rmi\pi s}{2}}\Li_{s}(\rme^{2\rmi\pi q})
+\rme^{\frac{\rmi\pi s}{2}}\Li_{s}(\rme^{-2\rmi\pi q})
\Big)\;,
\end{equation}
which is valid for $s\in\mathbb{C}\backslash\{0\}$ and $0<\Re\,q<1$. It corresponds to the constraint $-1/2<\Im\,c<1/2$ for $\eta(c)$. In order to go beyond this interval, one has to make the analytic continuation across the cut $[1,\infty[$ of the polylogarithm. This can be done by adding or removing a few terms from the sum in the definition (\ref{zeta def}): for $m\in\mathbb{Z}$
\begin{equation}
\label{difference zeta}
\zeta(s,q)=\zeta(s,q+m)+
\Bigg\{
\begin{array}{lc}
+\sum_{j=0}^{m-1}(q+j)^{-s} & \text{if}\,m>0\\
-\sum_{j=1}^{-m}(q-j)^{-s} & \text{if}\,m<0
\end{array}\;.
\end{equation}
One finds
\begin{eqnarray}
\label{eta[A,Li]}
\eta(c)=\Li_{5/2}(-\rme^{2\pi c})\\
\hspace{13mm}
-\frac{8\sqrt{2}\,\rmi\pi^{2}}{3}\Bigg(\sum_{a\in A_{0}^{+}}(c+\rmi a)^{3/2}+\sum_{a\in A^{-}}(c+\rmi a)^{3/2}\nonumber\\
\hspace{40mm}
-\sum_{a\in A_{0}^{-}}(c-\rmi a)^{3/2}-\sum_{a\in A^{+}}(c-\rmi a)^{3/2}\nonumber\\
\hspace{40mm}
+2\,\sgn(\Im\,c)\sum_{j=1}^{|[\Im\,c]|}\Big(c-\sgn(\Im\,c)\rmi(j-\tfrac{1}{2})\Big)^{3/2}\Bigg)\;,\nonumber
\end{eqnarray}
where $[x]$ is the integer closest to $x$ and $\sgn$ the sign function.
\end{subsection}

\begin{subsection}{Expression with a regularized infinite sum}
It is possible to remove completely Hurwitz zeta function from the definition (\ref{eta[A,zeta]}) of $\eta(c)$ by using (\ref{difference zeta}). For large $M$, the quantity $\zeta(s,q+M)$ has the asymptotics (\ref{zeta asymptotic}), and one finds for $\Re\,c>0$
\begin{eqnarray}
\label{eta[A] infinite sum}
\eta(c)=\lim_{M\to\infty}\frac{16\pi^{2}}{3}\,\Big(\frac{2}{5}\,M^{5/2}+c M^{3/2}-\frac{1+12c^{2}}{16}\,\sqrt{M}\Big)\\
-\frac{8\sqrt{2}\,\rmi\pi^{2}}{3}\,\Bigg(\sum_{a\in A_{0}^{+}}(c+\rmi a)^{3/2}+\sum_{a\in A^{-}}(c+\rmi a)^{3/2}+\sum_{j=0}^{M-1}\Big(c+\rmi(-j-\tfrac{1}{2})\Big)^{3/2}\nonumber\\
\hspace{18mm}
-\sum_{a\in A_{0}^{-}}(c-\rmi a)^{3/2}-\sum_{a\in A^{+}}(c-\rmi a)^{3/2}-\sum_{j=0}^{M-1}\Big(c-\rmi(-j-\tfrac{1}{2})\Big)^{3/2}\Bigg)\;.\nonumber
\end{eqnarray}
This expression corresponds to adding an infinity of negative half-integers to the sets $A_{0}^{+}$ and $A^{+}$, or to the sets $A_{0}^{-}$ and $A^{-}$.
\end{subsection}

\begin{subsection}{Approximate expression for large argument}
\label{subsection approximation eta}
For large $|c|$, one can use for $\eta$ the asymptotics (\ref{zeta asymptotic}) with $s=5/2$: $\zeta(-3/2,z)\simeq-2z^{5/2}/5+z^{3/2}/2-\sqrt{z}/8$ for $|z|\to\infty$. One has
\begin{eqnarray}
\label{eta[A] approx}
\eta(c)\simeq
-\frac{32\sqrt{2}\pi^{2}}{15}\,c^{5/2}
-\frac{\sqrt{2}\pi^{2}}{3}\,\sqrt{c}\\
\hspace{13mm}
-\frac{8\sqrt{2}\,\rmi\pi^{2}}{3}\Big(\sum_{a\in A_{0}^{+}}(c+\rmi a)^{3/2}+\sum_{a\in A^{-}}(c+\rmi a)^{3/2}\nonumber\\
\hspace{55mm}
-\sum_{a\in A_{0}^{-}}(c-\rmi a)^{3/2}-\sum_{a\in A^{+}}(c-\rmi a)^{3/2}\Big)\;.\nonumber
\end{eqnarray}
The equation for $c$ (\ref{eq eta' c}) with $\eta$ replaced by its approximate value for large $|c|$ has the form
\begin{equation}
\label{eq eta tilde' c}
\frac{4\rmi}{3}\,c^{3/2}\simeq\sum_{j=1}^{m}h_{p_{j},q_{j}}(c)\;,
\end{equation}
where the numbers $p_{j}$ and $q_{j}$ are positive half-integers and $h_{p,q}(c)=\sqrt{c+\rmi p}-\sqrt{c-\rmi q}$.

The function $h_{p,q}$ is a continuous bijection from the domain $\mathcal{D}=\{c\in\mathbb{C},\,\Re\,c>0\}$ to its image by $h_{p,q}$. Indeed, if we assume the existence of $c$ and $c'$ such that $h_{p,q}(c)=h_{p,q}(c')$, taking the square of $\sqrt{c+\rmi p}+\sqrt{c'-\rmi q}=\sqrt{c'+\rmi p}+\sqrt{c-\rmi q}$, simplifying, and taking again the square leads to $\rmi(p+q)(c-c')=0$, hence $c=c'$. The image by $h_{p,q}$ of $\mathcal{D}$ can thus be found by looking at the image of its boundary $\partial\mathcal{D}=\{c\in\mathbb{C},\,\Re\,c=0\}$. Writing $c=\rmi d$, $d\in\mathbb{R}$, one has for $p,q>0$
\begin{equation}
h_{p,q}(\rmi d)=
\left\{
\begin{array}{llc}
\rme^{3\rmi\pi/4}(\sqrt{q-d}-\sqrt{-p-d}) && \hspace{14.5mm} d<-p\\
\rme^{\rmi\pi/4}\sqrt{d+p}+\rme^{3\rmi\pi/4}\sqrt{q-d} && -p<d<q\\
\rme^{\rmi\pi/4}(\sqrt{d+p}-\sqrt{d-q}) && q<d \hspace{4.5mm}
\end{array}
\right.
\;.
\end{equation}
We note that $h_{p,q}(\rmi d)\to0$ when $d\to\pm\infty$, hence $h_{p,q}(\partial\mathcal{D})$ is a closed contour. The image by $h_{p,q}$ of $\mathcal{D}$ is the inside of this contour. All the elements $z\in h_{p,q}(\mathcal{D})$ are thus contained in the wedge $\frac{\pi}{4}<\arg z<\frac{3\pi}{4}$. But the sum of two points from this wedge still belongs to the same wedge. It implies that the right hand side of (\ref{eq eta tilde' c}) is also inside the wedge if $\Re\,c>0$. The equality (\ref{eq eta tilde' c}) then gives $\frac{\pi}{4}<\arg(\rmi c^{3/2})<\frac{3\pi}{4}$. This is only possible if $-\frac{\pi}{6}<\arg c<\frac{\pi}{6}$: all the solutions in $\mathcal{D}$ of the approximate equation (\ref{eq eta tilde' c}) must verify this stronger constraint.

We conjecture that $-\frac{\pi}{6}<\arg c<\frac{\pi}{6}$ is also verified for all the solutions of the exact equation (\ref{eq eta' c}) for $c$. Indeed, one observes in table \ref{table first eta c} that the constraint is verified for the first eigenstates, which are the ones where (\ref{eq eta' c}) differs the most from (\ref{eq eta tilde' c}) since they correspond to smaller values of $|c|$.
\end{subsection}

\begin{subsection}{Geometric picture of \texorpdfstring{$\eta'$}{eta'}}
The function $\eta'$, which appears in the equation (\ref{eq eta' c}) for $c$, involves sums of terms of the form $w=\pm\sqrt{c\pm\rmi a}$ with $a\in\mathbb{N}+1/2$.

Writing $w=x+\rmi y$ with $x,y\in\mathbb{R}$ and taking the real part of $w^{2}$ gives the equation $x^{2}-y^{2}=\Re\,c$. This is the equation of an hyperbola with foci at $\pm\sqrt{2\,\Re\,c}$ and perpendicular asymptotes. In the domain $\Re\,c>0$, the terms $w=\sqrt{c+\rmi a}$ belong to the branch of the hyperbola with $\Re\,w>0$, while the terms $w=-\sqrt{c-\rmi a}$ belong to the branch of the hyperbola with $\Re\,w<0$. The condition (\ref{m+-}) on the sets $A_{0}^{\pm}$ and $A^{\pm}$ implies that there is the same number of points on each branch of the hyperbola, although two points can be at the same location, like in figure \ref{fig hyperbola Cassini}.

Furthermore, from $w=\pm\sqrt{c\pm\rmi a}$, one has also $|w+\sqrt{c}|\,|w-\sqrt{c}|=a$. This is the equation of a Cassini oval with parameter $a\in\mathbb{N}+1/2$ and foci $\pm\sqrt{c}$ located on the hyperbola.

Considering the expression (\ref{eta[A] infinite sum}) for $\eta'$, one can also add to the picture of figure \ref{fig hyperbola Cassini} a number $M\to\infty$ of points on each branch of the hyperbola, in the region with negative imaginary part. We did not manage, however, to find a natural geometrical interpretation for the regularization term $M^{3/2}-\frac{3}{2}\sqrt{M}c\simeq(M-c)^{3/2}$.
\begin{figure}
  \begin{center}
    \begin{tabular}{c}
      \includegraphics[width=100mm]{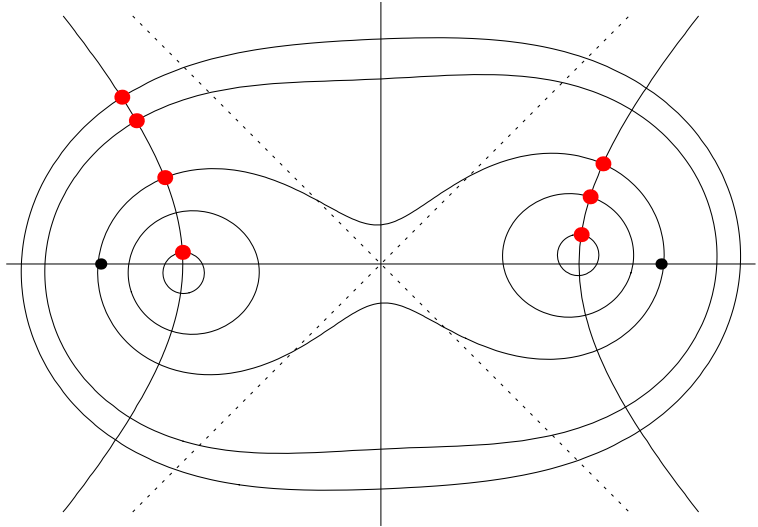}
      \end{tabular}
  \end{center}
  \caption{Geometric representation in the complex plane of the terms of the $4$ summations in expression (\ref{eta[A,zeta]}) for $\eta'(c)=0$. The eigenstate chosen is indexed by the sets $A_{0}^{+}=\{\frac{1}{2},\frac{3}{2},\frac{5}{2}\}$, $A^{-}=\{\frac{1}{2}\}$, $A_{0}^{-}=\{\frac{9}{2}\}$, $A^{+}=\{\frac{1}{2},\frac{5}{2},\frac{11}{2}\}$, and one has $c\approx2.397+0.218\,\rmi$. The red points, located at intersections between a hyperbola and Cassini ovals, represent the complex numbers $-\sqrt{c-\rmi a}$, $a\in A_{0}^{-}\cup A^{+}$ and $\sqrt{c+\rmi a}$, $a\in A_{0}^{+}\cup A^{-}$. The two smaller, black dots located on the horizontal axis at $\pm\sqrt{2\,\Re\,c}$ are the foci of the hyperbola, and the two perpendicular dotted lines its asymptotes.}
  \label{fig hyperbola Cassini}
\end{figure}
\end{subsection}
\end{section}

%%%%%%%%%%%%%%%%%
%%             %%
%%  Section 5  %%
%%             %%
%%%%%%%%%%%%%%%%%
\begin{section}{Numerical evaluations}
\label{section numerics}
The parameter $c$ solution of (\ref{eq eta' c}) and the quantity $\eta(c)$ defined in (\ref{eta[A,zeta]}) and related to the eigenvalues of the Markov matrix by (\ref{E[c]}), are evaluated numerically for the first eigenstates in this section.

\begin{subsection}{First excited state}
The two excited states whose eigenvalues have the smallest real part (gap) correspond to either $A_{0}^{+}=A^{+}=\{\frac{1}{2}\}$, $A_{0}^{-}=A^{-}=\{\}$ or $A_{0}^{+}=A^{+}=\{\}$, $A_{0}^{-}=A^{-}=\{\frac{1}{2}\}$. In both cases, the solution of (\ref{eq eta' c}) is $c=0.559534\ldots\in\mathbb{R}$. From (\ref{eta[A,zeta]}), this value of $c$ is solution of $\Li_{3/2}(-\rme^{2\rmi\pi c})=2\rmi\pi(\sqrt{2c+\rmi}-\sqrt{2c-\rmi})$, which is the same as equation (27) of \cite{GM2005.1} with $u=c/2$. Then, the corresponding eigenvalue of the Markov matrix is (\ref{E[c]}) with $P=\pm1$ and $\eta(c)=\Li_{5/2}(-\rme^{2\pi c})-\frac{4\rmi\pi^{2}}{3}((2c+\rmi)^{3/2}-(2c-\rmi)^{3/2})=13.0184\ldots$. This matches with equation (30) of \cite{GM2005.1}.
\end{subsection}

\begin{subsection}{Higher excited states}
The equation (\ref{eq eta' c}) for $c$ can be solved numerically for eigenstates beyond the gap. In figure \ref{fig first eta c}, $\eta(c)$ and $c$ are plotted for all the eigenstates whose index $Q$, defined in (\ref{Q[A]}), is between $1$ and $7$. The numerical values of $\eta(c)$ and $c$ are also given in table \ref{table first eta c} for all the eigenstates with $Q$ between $0$ and $5$. The values of $\eta(c)$ for all the eigenstates with given $Q$ accumulate for large $Q$ in a crescent shape, see figure \ref{fig crescent Q=30}.
\begin{figure}
  \begin{center}
    \begin{tabular}{cc}
      \begin{tabular}{c}
      \includegraphics[width=70mm]{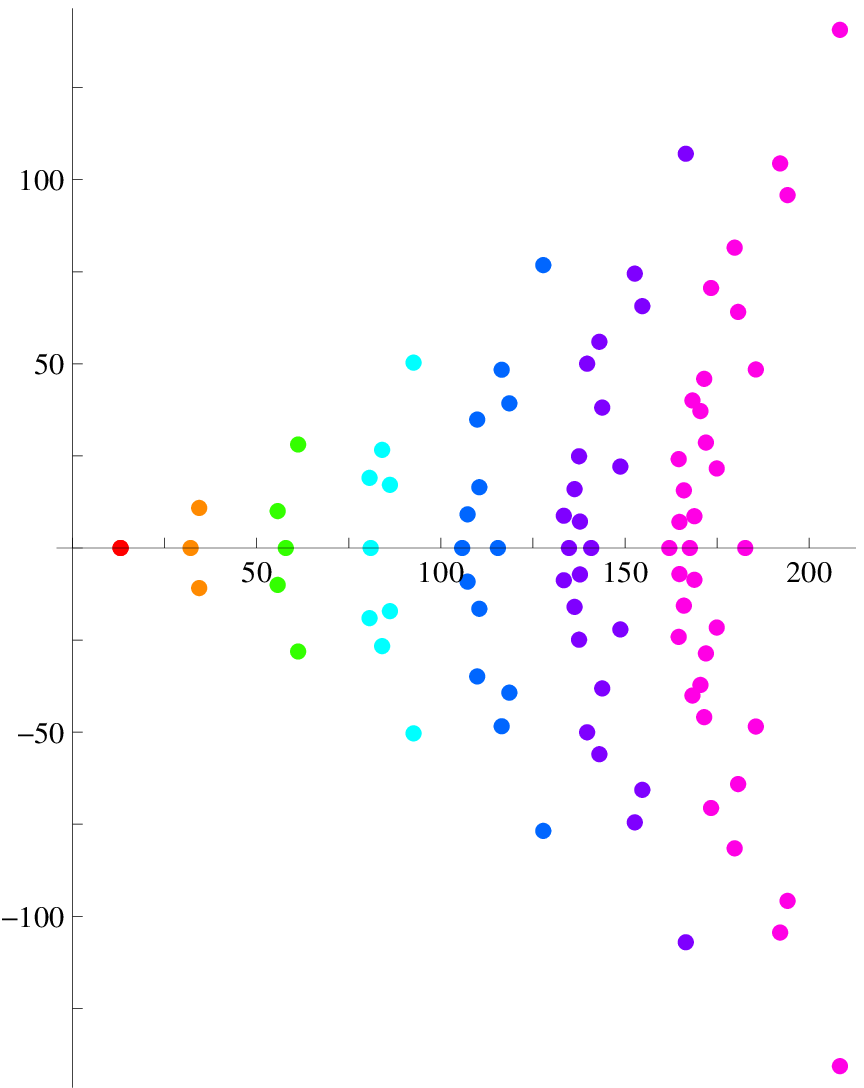}
      \end{tabular}
      &
      \begin{tabular}{c}
      \includegraphics[width=70mm]{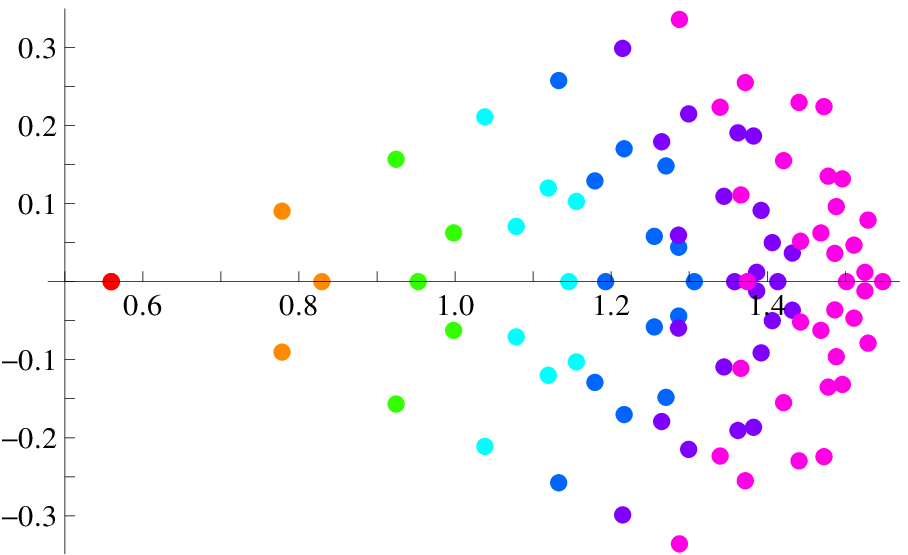}
      \end{tabular}
    \end{tabular}
  \end{center}
  \caption{On the left, values of $\eta(c)/\sqrt{2\pi}$ for the first eigenstates, with $\eta$ defined in (\ref{eta[A,zeta]}). The corresponding eigenvalues are given in terms of $\eta(c)$ in (\ref{E[c]}). On the right, intermediate parameter $c$ solution of $\eta'(c)=0$ and used in the calculation of the eigenvalues. The dots are coloured according to the value of the index $Q$, defined in (\ref{Q[A]}), which increases from left to right on both graphs. All the eigenstates with $Q$ between $1$ and $7$ are represented.}
  \label{fig first eta c}
\end{figure}
\begin{table}
  \begin{center}
  \begin{tabular}{cccccc}
  Q & $(A_{0}^{+},A^{-})$ & $(A_{0}^{-},A^{+})$ & Degeneracy & $c$ & $\displaystyle\frac{\eta(c)}{\sqrt{2\pi}}$\\
  0 & $()$ & $()$ & 1 & Indeterminate & Indeterminate\\\hline
  1 & $(\frac{1}{2})$ & $(\frac{1}{2})$ & 2 & $0.559534$ & $13.0184$\\\hline
  2 & $(\frac{1}{2},\frac{1}{2})$ & $(\frac{1}{2},\frac{1}{2})$ & 1 & $0.82937$ & $32.0353$\\
  2 & $(\frac{1}{2})$ & $(\frac{3}{2})$ & 2 & $0.778306+0.0902648\,\rmi$ & $34.3768-10.8732\,\rmi$\\
  2 & $(\frac{3}{2})$ & $(\frac{1}{2})$ & 2 & $0.778306-0.0902648\,\rmi$ & $34.3768+10.8732\,\rmi$\\\hline
  3 & $(\frac{1}{2},\frac{1}{2})$ & $(\frac{1}{2},\frac{3}{2})$ & 2 & $0.998165+0.0624517\,\rmi$ & $55.6906-10.0233\,\rmi$\\
  3 & $(\frac{1}{2},\frac{3}{2})$ & $(\frac{1}{2},\frac{1}{2})$ & 2 & $0.998165-0.0624517\,\rmi$ & $55.6906+10.0233\,\rmi$\\
  3 & $(\frac{3}{2})$ & $(\frac{3}{2})$ & 2 & $0.952726$ & $57.8869$\\
  3 & $(\frac{1}{2})$ & $(\frac{5}{2})$ & 2 & $0.924474+0.156784\,\rmi$ & $61.2404-28.1045\,\rmi$\\
  3 & $(\frac{5}{2})$ & $(\frac{1}{2})$ & 2 & $0.924474-0.156784\,\rmi$ & $61.2404+28.1045\,\rmi$\\\hline
  4 & $(\frac{1}{2},\frac{1}{2})$ & $(\frac{3}{2},\frac{3}{2})$ & 1 & $1.15545+0.102772\,\rmi$ & $80.6252-19.0466\,\rmi$\\
  4 & $(\frac{3}{2},\frac{3}{2})$ & $(\frac{1}{2},\frac{1}{2})$ & 1 & $1.15545-0.102772\,\rmi$ & $80.6252+19.0466\,\rmi$\\
  4 & $(\frac{1}{2},\frac{3}{2})$ & $(\frac{1}{2},\frac{3}{2})$ & 6 & $1.14571$ & $80.9397$\\
  4 & $(\frac{1}{2},\frac{1}{2})$ & $(\frac{1}{2},\frac{5}{2})$ & 2 & $1.11938+0.119929\,\rmi$ & $84.0436-26.6439\,\rmi$\\
  4 & $(\frac{1}{2},\frac{5}{2})$ & $(\frac{1}{2},\frac{1}{2})$ & 2 & $1.11938-0.119929\,\rmi$ & $84.0436+26.6439\,\rmi$\\
  4 & $(\frac{3}{2})$ & $(\frac{5}{2})$ & 2 & $1.07836+0.0707095\,\rmi$ & $86.1719-17.1468\,\rmi$\\
  4 & $(\frac{5}{2})$ & $(\frac{3}{2})$ & 2 & $1.07836-0.0707095\,\rmi$ & $86.1719+17.1468\,\rmi$\\
  4 & $(\frac{1}{2})$ & $(\frac{7}{2})$ & 2 & $1.03814+0.211083\,\rmi$ & $92.5825-50.3354\,\rmi$\\
  4 & $(\frac{7}{2})$ & $(\frac{1}{2})$ & 2 & $1.03814-0.211083\,\rmi$ & $92.5825+50.3354\,\rmi$\\\hline
  5 & $(\frac{1}{2},\frac{1}{2},\frac{3}{2})$ & $(\frac{1}{2},\frac{1}{2},\frac{3}{2})$ & 2 & $1.30655$ & $105.785$\\
  5 & $(\frac{1}{2},\frac{3}{2})$ & $(\frac{3}{2},\frac{3}{2})$ & 2 & $1.28648+0.0440786\,\rmi$ & $107.25-9.1338\,\rmi$\\
  5 & $(\frac{3}{2},\frac{3}{2})$ & $(\frac{1}{2},\frac{3}{2})$ & 2 & $1.28648\, -0.0440786\,\rmi$ & $107.25+9.1338\,\rmi$\\
  5 & $(\frac{1}{2},\frac{1}{2})$ & $(\frac{3}{2},\frac{5}{2})$ & 2 & $1.27024\, +0.14822\,\rmi$ & $109.865-34.8779\,\rmi$\\
  5 & $(\frac{3}{2},\frac{5}{2})$ & $(\frac{1}{2},\frac{1}{2})$ & 2 & $1.27024\, -0.14822\,\rmi$ & $109.865+34.8779\,\rmi$\\
  5 & $(\frac{1}{2},\frac{3}{2})$ & $(\frac{1}{2},\frac{5}{2})$ & 6 & $1.25528\, +0.0580091\,\rmi$ & $110.44-16.5029\,\rmi$\\
  5 & $(\frac{1}{2},\frac{5}{2})$ & $(\frac{1}{2},\frac{3}{2})$ & 6 & $1.25528\, -0.0580091\,\rmi$ & $110.44+16.5029\,\rmi$\\
  5 & $(\frac{5}{2})$ & $(\frac{5}{2})$ & 2 & $1.19288$ & $115.49$\\
  5 & $(\frac{1}{2},\frac{1}{2})$ & $(\frac{1}{2},\frac{7}{2})$ & 2 & $1.21666\, +0.170272\,\rmi$ & $116.513-48.4036\,\rmi$\\
  5 & $(\frac{1}{2},\frac{7}{2})$ & $(\frac{1}{2},\frac{1}{2})$ & 2 & $1.21666\, -0.170272\,\rmi$ & $116.513+48.4036\,\rmi$\\
  5 & $(\frac{3}{2})$ & $(\frac{7}{2})$ & 2 & $1.17919\, +0.129087\,\rmi$ & $118.594-39.2823\,\rmi$\\
  5 & $(\frac{7}{2})$ & $(\frac{3}{2})$ & 2 & $1.17919\, -0.129087\,\rmi$ & $118.594+39.2823\,\rmi$\\
  5 & $(\frac{1}{2})$ & $(\frac{9}{2})$ & 2 & $1.13278\, +0.257656\,\rmi$ & $127.805-76.8167\,\rmi$\\
  5 & $(\frac{9}{2})$ & $(\frac{1}{2})$ & 2 & $1.13278\, -0.257656\,\rmi$ & $127.805+76.8167\,\rmi$\\
  \end{tabular}
  \end{center}
  \caption{Values of $c$ and $\eta(c)/\sqrt{2\pi}$ for the first eigenstates, with $\eta$ defined in (\ref{eta[A,zeta]}). The corresponding eigenvalues are given in terms of $\eta(c)$ in (\ref{E[c]}). The intermediate parameter $c$ solution of $\eta'(c)=0$ is used in the calculation of the eigenvalues. The index $Q$ in the first column is defined in (\ref{Q[A]}). All the eigenstates with $Q$ between $0$ and $5$ are given in the table. The columns $2$ and $3$ give sorted lists of the elements of $A_{0}^{\pm}$ and $A^{\mp}$ put together. An element contained in both $A_{0}^{\pm}$ and $A^{\mp}$ is given twice in the list. Column $4$ gives the number of eigenstates with the same value of $\eta(c)$.}
  \label{table first eta c}
\end{table}
\begin{figure}
  \begin{center}
    \begin{tabular}{c}
      \begin{picture}(0,0)
        \put(101,6){$L$}
        \put(-18,61){$-2L^{3/2}\,\Re\,E$}
      \end{picture}
      \includegraphics[width=100mm]{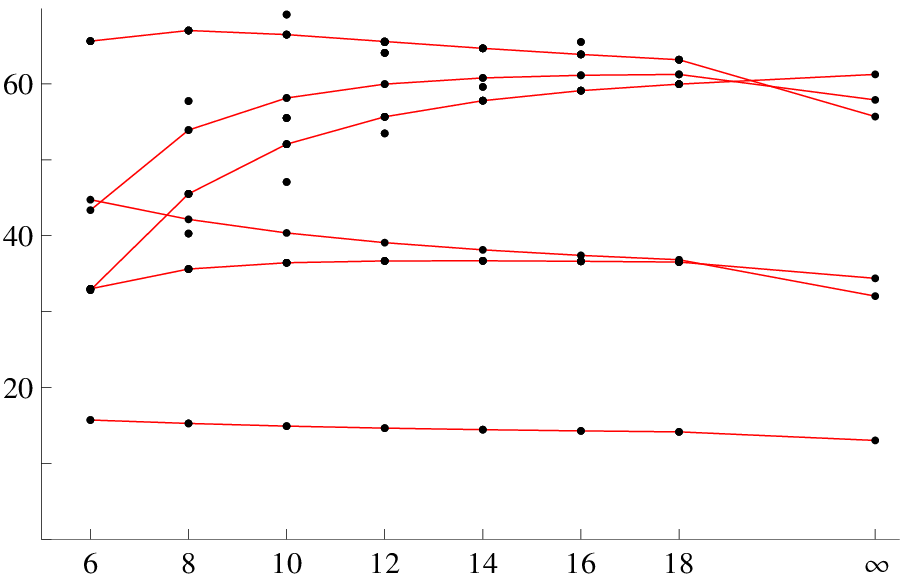}
      \end{tabular}
  \end{center}
  \caption{Bottom of the spectrum of TASEP for small system size $L$ with $N=L/2$ particles (density $\rho=1/2$). The black dots correspond to $-2L^{3/2}\,\Re\,E$ for each eigenvalue $E$. The points for finite $L$ are obtained from exact diagonalization of the Markov matrix $M$. The rightmost points represent the $L\to\infty$ limit given by (\ref{E[c]}). The eigenvalues linked correspond to a similar choice of integers $k_{j}$ in the solution (\ref{y[k,b]}) of the Bethe ansatz equations.}
  \label{fig spectrum small L}
\end{figure}

We observe in table \ref{table first eta c} that the eigenstates forming the inner part (where $\Re\,\eta(c)$ is larger) of the crescent correspond to either $A_{0}^{+}=\{a\}$, $A^{+}=\{Q-a\}$, $A_{0}^{-}=A^{-}=\{\}$ or to $A_{0}^{-}=\{Q-a\}$, $A^{-}=\{a\}$, $A_{0}^{+}=A^{+}=\{\}$, with $a$ half integer between $\frac{1}{2}$ and $Q-\frac{1}{2}$. At large $Q$ with $\alpha=a/Q$ fixed, (\ref{eq eta' c}) and (\ref{eta[A] approx}) imply
\begin{eqnarray}
\label{crescent inner}
c\simeq\frac{3^{2/3}}{2^{4/3}}\,(\rme^{-\rmi\pi/4}\sqrt{\alpha}+\rme^{\rmi\pi/4}\sqrt{1-\alpha})^{2/3}Q^{1/3}\\
\eta(c)\simeq\frac{8\sqrt{2}\pi^{2}}{3}\,(\rme^{\rmi\pi/4}\alpha^{3/2}+\rme^{-\rmi\pi/4}(1-\alpha)^{3/2})Q^{3/2}\;.
\end{eqnarray}
The first equation indicates that the parameter $c$ belongs to the arc of circle $|c|\simeq(3/4)^{2/3}Q^{1/3}$, $-\frac{\pi}{6}<\arg c<\frac{\pi}{6}$. The second equation shows that $-\frac{\pi}{4}<\arg\eta(c)<\frac{\pi}{4}$ for large $Q$.
\begin{figure}
  \begin{center}
    \begin{tabular}{c}
      \includegraphics[width=50mm]{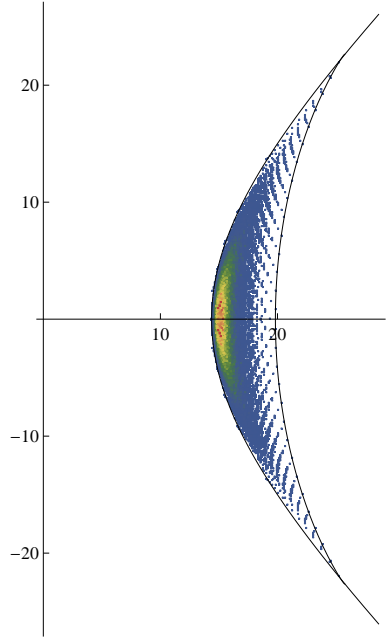}
      \end{tabular}
  \end{center}
  \caption{Values in the complex plane of $\eta(c)/(\sqrt{2\pi}Q^{5/4})$ for all $589128$ eigenstates corresponding to an index $Q=30$, with $\eta$ defined in (\ref{eta[A,zeta]}) and $Q$ defined in (\ref{Q[A]}). The corresponding eigenvalues are given in terms of $\eta(c)$ in (\ref{E[c]}). Each small square is coloured according to the number of eigenstates within the square, with smaller numbers at the border of the crescent. The curve at the inner edge of the crescent (right) is given by the exact value of $\eta(c)$ with the sets given before (\ref{crescent inner}) with $\alpha\in[0,1]$. The curve at the outer edge of the crescent (left) is given by the asymptotics (\ref{crescent outer}) with $1/10<\alpha<10$.}
  \label{fig crescent Q=30}
\end{figure}

All eigenstates in the bulk of the crescent (except for a negligible fraction of them when $Q\to\infty$) correspond to sets $A_{0}^{\pm}$, $A^{\pm}$ with typical elements scaling as $\sqrt{Q}$ and of cardinal $m_{\pm}\simeq\mu_{\pm}\sqrt{Q}$. We define for $Q^{-1/2}\ll\rmd u\ll1$ smooth functions $\theta$ and $\overline{\theta}$ such that $2\theta(u)\sqrt{Q}\,\rmd u$ (respectively $2\overline{\theta}(u)\sqrt{Q}\,\rmd u$) is the number of elements of $A_{0}^{+}$ (resp. $A_{0}^{-}$) plus the number of elements of $A^{-}$ (resp. $A^{+}$) in the interval $\sqrt{Q}[u,u+du]$. They verify the normalizations $\int_{0}^{\infty}\rmd u\,\theta(u)=\int_{0}^{\infty}\rmd u\,\overline{\theta}(u)=\mu_{+}+\mu_{-}$ and $\int_{0}^{\infty}\rmd u\,u\,\theta(u)=\int_{0}^{\infty}\rmd u\,u\,\overline{\theta}(u)=1$. Writing $c=\chi\sqrt{Q}$, (\ref{eta[A] approx}) gives for large $Q$
\begin{equation}
\label{eta[theta]}
\fl\hspace{5mm}
\frac{\eta(c)}{Q^{5/4}}\simeq
-\frac{32\sqrt{2}\pi^{2}}{15}\,\chi^{5/2}-\frac{8\sqrt{2}\,\rmi\pi^{2}}{3}\,\int_{0}^{\infty}\rmd u\,\Big(\theta(u)(\chi+\rmi u)^{3/2}-\overline{\theta}(u)(\chi-\rmi u)^{3/2}\Big)\;.
\end{equation}
The total number of ways to choose the elements of the sets $A_{0}^{+}$ and $A^{-}$ in the interval $[u,u+\rmd u]$ is equal to
\begin{equation}
\omega(u)\simeq\int_{0}^{2\theta(u)}\rmd\sigma\,\C{\sqrt{Q}\,\rmd u}{\sigma\sqrt{Q}\,\rmd u}\C{\sqrt{Q}\,\rmd u}{(2\theta(u)-\sigma)\sqrt{Q}\,\rmd u}\simeq\rme^{2\sqrt{Q}\,s(\theta(u))\,\rmd u}\;,
\end{equation}
with $s(\varphi)=-\varphi\log\varphi-(1-\varphi)\log(1-\varphi)$. Similarly for $A_{0}^{-}$ and $A^{+}$, the number of choices is $\overline{\omega}(u)\simeq\exp[2\sqrt{Q}\,s(\overline{\theta}(u))\,\rmd u]$. The total number of eigenstates corresponding to $\theta$ and $\overline{\theta}$ is then equal to $\Omega[\theta,\overline{\theta}]\simeq\exp[2\sqrt{Q}\int_{0}^{\infty}\rmd u\,(s(\theta(u))+s(\overline{\theta}(u)))]$.

The outer part (where $\Re\,\eta(c)$ is smaller) of the crescent in figure \ref{fig first eta c} is reached from the bulk of the crescent for functions $\theta$ and $\overline{\theta}$ such that $\log\Omega[\theta,\overline{\theta}]$ vanishes. It corresponds to $\theta(u)$ and $\overline{\theta}(u)$ equal to either $0$ or $1$, \textit{i.e.} the sets $A_{0}^{+}=A^{-}$ and $A_{0}^{-}=A^{+}$ are both composed by a finite reunion of intervals of the form $\{a,a+1,\ldots,b\}$ with $b-a$ scaling as $\sqrt{Q}$. For simplicity, we consider only the case of sets equal to only one interval: we write $A_{0}^{+}=A^{-}=[\sqrt{Q}\,x,\sqrt{Q}\,y]$ and $A_{0}^{-}=A^{+}=[\sqrt{Q}\,\overline{x},\sqrt{Q}\,\overline{y}]$ and call $\mu=\mu_{+}=\mu_{-}$. The normalization conditions for $\theta$ and $\overline{\theta}$ imply $y-x=\overline{y}-\overline{x}=\mu$ and $(y^{2}-x^{2})+(\overline{y}^{2}-\overline{x}^{2})=1$. Writing $\frac{x+y}{2}=\frac{1}{4\mu}+\lambda$ and $\frac{\overline{x}+\overline{y}}{2}=\frac{1}{4\mu}-\lambda$, the positivity of $x$ and $\overline{x}$ implies $\frac{\mu}{2}-\frac{1}{4\mu}\leq\lambda\leq\frac{1}{4\mu}-\frac{\mu}{2}$, thus $0\leq\mu\leq1/\sqrt{2}$. The quantity $Q^{-5/4}\eta(\chi\sqrt{Q})$ depends only on $\chi$ and on the parameters $\lambda$ and $\mu$. We observe numerically that the values of $\eta$ with smallest real part correspond to the two extremal choices for $\lambda$: $\lambda=\frac{\mu}{2}-\frac{1}{4\mu}$ and $\lambda=\frac{1}{4\mu}-\frac{\mu}{2}$. Since the two values of $\lambda$ are exchanged by $\mu\to(2\mu)^{-1}$, it is sufficient to consider only one of them and take $\mu\in\mathbb{R}^{+}$. A parametric expression with parameter $\alpha=\sqrt{2}\mu$ for the outer part of the crescent is then given by
\begin{equation}
\label{crescent outer}
\fl\hspace{10mm}
\frac{\eta(c)}{Q^{5/4}}=-\frac{32\sqrt{2}\pi^{2}}{15}\Big[\Big(\chi+\frac{\rmi\alpha}{\sqrt{2}}\Big)^{5/2}+\Big(\chi-\frac{\rmi\alpha^{-1}}{\sqrt{2}}\Big)^{5/2}-\Big(\chi+\frac{\rmi(\alpha-\alpha^{-1})}{\sqrt{2}}\Big)^{5/2}\Big]\;
\end{equation}
with $c=\chi\sqrt{Q}$ solution of (\ref{eq eta' c}). When $\alpha\to0$, the solution is $c\simeq2^{-7/6}3^{2/3}\rme^{\rmi\pi/6}\alpha^{1/3}\sqrt{Q}$ and $\eta(c)\simeq2^{13/4}3^{-1}\pi^{2}\rme^{-\rmi\pi/4}\alpha^{-1/2}Q^{5/4}$. The solution for $\alpha\to\infty$ follows from $\alpha\to\alpha^{-1}$ and complex conjugation. We find again that the edges of the crescent verify $\arg c\to\pm\pi/6$ and $\arg\eta(c)\to\mp\pi/4$. The point of the crescent with smallest real part corresponds to $\alpha=1$, for which $\eta(c)/\sqrt{2\pi}=32\pi^{3/2}z^{1/4}/15\approx14.376662$ with $z$ the largest real root of $395307-642600z+212544z^{2}+512z^{3}=0$.
\begin{table}
  \begin{center}
  \begin{tabular}{cccccc}
  & $(\frac{1}{2})(\frac{1}{2})$ & \makebox[0mm]{$(\frac{1}{2},\frac{1}{2})(\frac{1}{2},\frac{1}{2})$} & $(\frac{1}{2})(\frac{3}{2})$ & $(\frac{1}{2},\frac{1}{2})(\frac{1}{2},\frac{3}{2})$ & $(\frac{3}{2})(\frac{3}{2})$\\
  &&&&&\\
  $L=6$ & $15.7089$ & $44.7384$ & $32.9964-21.8944\,\rmi$ & $65.6302-23.2962\,\rmi$ & $43.3573$\\
  $L=8$ & $15.2587$ & $42.1495$ & $35.6021-20.2845\,\rmi$ & $67.0315-21.2037\,\rmi$ & $53.9139$\\
  $L=10$ & $14.9056$ & $40.3553$ & $36.4302-18.8088\,\rmi$ & $66.4897-19.3528\,\rmi$ & $58.1394$\\
  $L=12$ & $14.6395$ & $39.0764$ & $36.6691-17.6584\,\rmi$ & $65.5817-17.9374\,\rmi$ & $59.9675$\\
  $L=14$ & $14.4357$ & $38.1283$ & $36.6883-16.7714\,\rmi$ & $64.682-16.8607\,\rmi$ & $60.7825$\\
  $L=16$ & $14.2759$ & $37.4006$ & $36.6216-16.077\,\rmi$ & $63.878-16.0264\,\rmi$ & $61.1294$\\
  $L=18$ & $14.1477$ & $36.8258$ & $36.5227-15.5226\,\rmi$ & $63.1801-15.3657\,\rmi$ & $61.2461$\\
  &&&&&\\
  BST & $13.0184$ & $32.0353$ & $34.3771-10.8731\,\rmi$ & $55.6919-10.0232\,\rmi$ & $57.8772$\\
  Exact result & $13.0184$ & $32.0353$ & $34.3768-10.8732\,\rmi$ & $55.6906-10.0233\,\rmi$ & $57.8869$
  \end{tabular}
  \end{center}
  \caption{Some eigenvalues of TASEP for small system size $L$ with $N=L/2$ particles (density $\rho=1/2$). Each column corresponds to a given choice of the sets $A_{0}^{\pm}$ and $A^{\mp}$ which index the first eigenstates in the Bethe ansatz formulation. At the top of each column are sorted lists of the elements of $A_{0}^{\pm}$ and $A^{\mp}$ put together. An element contained in both $A_{0}^{\pm}$ and $A^{\mp}$ is given twice in the list. The numbers correspond to $-2L^{3/2}\,\Re\,E$ for an eigenvalue $E$. The penultimate row gives the extrapolation of these numbers using BST algorithm with exponent $\omega=1$. The last line is the exact limit $L\to\infty$ given by (\ref{E[c]}).}
  \label{table first E BST}
\end{table}
\end{subsection}

\begin{subsection}{Numerical checks by exact diagonalization}
We checked by exact diagonalization for small $L$ the numerical values of table \ref{table first eta c} obtained from the large $L$ limit (\ref{E[c]}). We used BST algorithm (see \textit{e.g.} \cite{HS1988.1}) to extrapolate the data from $6\leq L\leq18$, $\rho=1/2$ up to $L\to\infty$. BST algorithm consists in fitting the numerical data with a Taylor series in $L^{-\omega}$. We have chosen $\omega=1$ in our computations.

The results, given in table \ref{table first E BST}, are in good agreement with the numerical values of table \ref{table first eta c}. A difficulty in doing the extrapolation is that one must choose for all $L$ the "same" eigenvalue, with numbers $k_{j}$ in (\ref{E[psi]}) and (\ref{b[psi]}) corresponding to the same sets $A_{0}^{\pm}$, $A^{\pm}$. Indeed, we observe in figure \ref{fig spectrum small L} that there are many crossings of eigenvalues as $L$ increases from small values to infinity. It means that Bethe ansatz is somewhat needed in order to perform the extrapolation on the numerical data from exact diagonalization.
\end{subsection}
\end{section}

%%%%%%%%%%%%%%%%%
%%             %%
%%  Section C  %%
%%             %%
%%%%%%%%%%%%%%%%%
\begin{section}{Conclusion}
Using Bethe ansatz, we studied the lower edge of the spectrum of the Markov matrix and of the transfer matrix of TASEP on a ring. In the lower part of the spectrum, eigenvalues have real part scaling as $L^{-3/2}$ when the system size $L$ goes to infinity with fixed density of particles. It corresponds to a relaxation time growing as $L^{z}$ with the dynamical exponent $z=3/2$ of the KPZ universality class.

We found that each eigenstate is characterized by a function $\eta$, defined in (\ref{eta[A,zeta]}): the corresponding eigenvalue of the Markov matrix is written (\ref{E[c]}) in terms of $\eta(c)$, where the complex number $c$ is determined by the saddle point equation $\eta'(c)=0$. This property extends to the eigenvalues of the transfer matrix (\ref{E[lambda,c] 1}), (\ref{E[lambda,c] 2}).

It would be interesting to generalize the present work to ASEP, where particles can hop in both directions with different rates $p$ and $q$. Especially interesting would be the crossover to KPZ scale $p-q\sim1/\sqrt{L}$ and the crossover to equilibrium scale $p-q\sim1/L$. The Bethe ansatz for general ASEP is however much more complicated than the one for TASEP, since the unknown parameter $b$ is replaced by a function, solution of a nonlinear integral equation which replaces the equation (\ref{b[psi]}) for the constant $b$. Currently, only the first terms in the expansion for large $(p-q)\sqrt{L}$ of the gap are known exactly for ASEP \cite{K1995.1}, see also \cite{NdN1995.1} for results by exact diagonalization of small systems.

Another possible direction of study would be to consider the eigenstates corresponding to the lowest eigenvalues for TASEP, which would be necessary in order to compute the crossover between transient and stationary fluctuations of the current. This is usually a difficult problem. It should be noted, however, that once the eigenvalues are known, the determination of the eigenvectors reduces to a linear problem. Moreover, since the first eigenvalues of the whole transfer matrix depend only on the quantity $\eta(c)$, it is tempting to believe that components of the eigenvectors can be expressed in terms of $\eta(c)$ in the thermodynamic limit.
%\begin{subsection}*{Acknowledgements}
%
%\end{subsection}
\end{section}

\appendix
%%%%%%%%%%%%%%%%%
%%             %%
%%  Section A  %%
%%             %%
%%%%%%%%%%%%%%%%%
\begin{section}{Some properties of the function \texorpdfstring{$\psi$}{psi}}
\label{appendix psi}
Let $g$ be the function from $\mathbb{C}\backslash\mathbb{R}^{-}$ to $\mathbb{C}\backslash(\rme^{\rmi\pi\rho-b_{0}}[1,\infty[\,\cup\,\rme^{-\rmi\pi\rho-b_{0}}[1,\infty[)$ defined by
\begin{equation}
\label{g}
g(y)=\frac{1-y}{y^{\rho}}\;,
\end{equation}
with $0<\rho<1$ and $b_{0}$ defined in (\ref{b0}). It was argued in Appendix A of \cite{P2013.1} that $g$ is a bijection. With the usual definition of the logarithm in $y^{\rho}=\rme^{\rho\log(y)}$, the interval $\mathbb{R}^{-}$ is a branch cut of $g$. The reciprocal function $g^{-1}$ also has a branch cut, equal to $(\,\rme^{\rmi\pi\rho-b_{0}}[1,\infty[\,)\,\cup\,(\,\rme^{-\rmi\pi\rho-b_{0}}[1,\infty[\,)$, which is the image by $g$ of its branch cut (\textit{i.e.} the limit $\epsilon\to0$, $\epsilon>0$ of the set $\{g(-y+\rmi\epsilon),y\in\mathbb{R}^{-}\}\cup\{g(-y-\rmi\epsilon),y\in\mathbb{R}^{-}\}$).

The function $\psi$, defined as the logarithm of $g^{-1}$, is the solution of the implicit equation (\ref{psi}) with the additional condition $-\pi<\Im\,\psi(z)<\pi$. It verifies in particular $\psi(0)=0$. The function $\psi$ has the same branch cuts as $g^{-1}$, since the image by $g^{-1}$ of the cut of $g^{-1}$ is the cut of $g$, $\mathbb{R}^{-}$, which is also the cut of the logarithm. The value of $\psi$ at the branch points $\rme^{\pm\rmi\pi\rho-b_{0}}$ is
\begin{equation}
\label{psi(exp(...))}
\psi(\rme^{\pm\rmi\pi\rho-b_{0}-\epsilon})\underset{\epsilon\to0}{\to}\log\Big(\frac{\rho}{1-\rho}\Big)\mp\rmi\pi
\quad
\text{with}\;
-\pi<\arg\epsilon<\pi\;.
\end{equation}
The expansion of $\psi$ near the branch points can be obtained by inserting $z=\rme^{\pm\rmi\pi\rho-b_{0}-\epsilon}$ and the expansion $\psi(z)=\log(\rho/(1-\rho))\mp\rmi\pi+\sum_{k=0}^{\infty}f_{k}\epsilon^{k/2}$ in (\ref{psi}) and solving for the $f_{k}$. For small $\epsilon$ with $-\pi<\arg\epsilon<\pi$, one finds
\begin{eqnarray}
\label{psi(exp(...-epsilon))}
\psi(\rme^{\pm\rmi\pi\rho-b_{0}-\epsilon})
=\log\Big(\frac{\rho}{1-\rho}\Big)\mp\rmi\pi
\pm\frac{\rmi\sqrt{2}\sqrt{\epsilon}}{\sqrt{\rho(1-\rho)}}
+\frac{(1-2\rho)\epsilon}{3\rho(1-\rho)}\nonumber\\
\hspace{30mm}
\mp\frac{\rmi(1-\rho+\rho^{2})\epsilon^{3/2}}{9\sqrt{2}(\rho(1-\rho))^{3/2}}-\frac{(1+\rho)(2-\rho)(1-2\rho)\epsilon^{2}}{135\rho^{2}(1-\rho)^{2}}\nonumber\\
\hspace{30mm}
\pm\frac{\rmi(1-\rho+\rho^{2})^{2}\epsilon^{5/2}}{540\sqrt{2}(\rho(1-\rho))^{5/2}}+\O{\epsilon^{3}}\;.
\end{eqnarray}

An interesting property of the function $\psi$ is that the derivative $\psi'(z)$ can be expressed as a function of $\psi(z)$ only. Indeed writing $z$ as a function of $\psi(z)$ using (\ref{psi}) and taking the derivative with respect to $z$, one finds
\begin{equation}
\label{psi'}
\psi'(z)=-\frac{\rme^{\rho\psi(z)}}{\rho+(1-\rho)\rme^{\psi(z)}}\;.
\end{equation}
This property is used in section \ref{section low spectrum} to compute integrals where the integrand depends on $\psi(z)$, as it allows to explicitly perform the change of variable $y=\psi(z)$.
\end{section}

%%%%%%%%%%%%%%%%%
%%             %%
%%  Section B  %%
%%             %%
%%%%%%%%%%%%%%%%%
\begin{section}{Euler Maclaurin formula with square root singularity}
\label{appendix Euler Maclaurin}
In this appendix, we derive the asymptotic expansion of
\begin{equation}
\label{S(L,M)}
S_{L,M}(d)=\sum_{j=1}^{M}f\Big(\frac{j+d}{L}\Big)\;
\end{equation}
for large $L$, with $d$ and $\mu=M/L$ taken fixed, in the case where $f(x)\sim\sqrt{x}$ for $x\to0$. More specifically, we consider $f(x)$ of the form
\begin{equation}
\label{f[f]}
f(x)=\sum_{k=0}^{\infty}f_{k}\,x^{k/2}\;.
\end{equation}
We require this expansion to be valid for all $x$ such that either $0\leq\arg x\leq\arg(d+1)$ or $\arg(d+1)\leq\arg x\leq0$ depending on whether $d$ is in the upper half or in the lower half of the complex plane.

We introduce the Hurwitz zeta function (see \textit{e.g.} \cite{DLMF}, Section 25.11)
\begin{equation}
\label{zeta def}
\zeta(s,q)=\sum_{j=0}^{\infty}(j+q)^{-s}\;.
\end{equation}
This definition converges when $\Re\,s>1$, $q\not\in-\mathbb{N}$. For fixed $q$, $\zeta(s,q)$ can be continued to a holomorphic function of $s$ in $\mathbb{C}\backslash\{1\}$, with a simple pole at $s=1$. Then, for all $s\neq1$, $\zeta(s,q)$ can be analytically continued to a holomorphic function of the variable $q$ in the domain $\mathbb{C}\backslash\mathbb{R}^{-}$. Hurwitz zeta function verifies the asymptotic expansion for large $|q|$, $-\pi<\arg\,q<\pi$, $s\neq0$
\begin{equation}
\label{zeta asymptotic}
\zeta(1-s,q)\underset{|q|\to\infty}{\simeq}-\frac{q^{s}}{s}\sum_{r=0}^{\infty}\C{s}{r}\frac{B_{r}}{q^{r}}\;,
\end{equation}
where the $B_{r}$ are Bernoulli numbers: $B_{0}=1$, $B_{1}=-1/2$, $B_{2}=1/6$, $B_{3}=0$, $B_{4}=-1/30$, \ldots\,, $B_{r}=0$ for $r$ odd $\geq3$.

The large $L$ asymptotics of the sum (\ref{S(L,M)}) can be obtained by writing it in terms of Hurwitz zeta function as
\begin{equation}
S_{L,\mu L}(d)=\sum_{k=0}^{\infty}\frac{f_{k}}{L^{k/2}}\Big[\zeta(-\tfrac{k}{2},\,d+1)-\zeta(-\tfrac{k}{2},\,\mu L+d+1)\Big]\;.
\end{equation}
Indeed, by analytic continuation in $s$, one can always use (\ref{zeta def}) to write $\zeta(s,q)-\zeta(s,q+M)$ as a finite sum even when $\Re\,s$ is not larger than $1$. Applying (\ref{zeta asymptotic}) for the second $\zeta$, one finds for $\mu>0$ the asymptotic expansion
\begin{eqnarray}
\label{S[B,zeta]}
\sum_{j=1}^{\mu L}f\Big(\frac{j+d}{L}\Big)\simeq L\Big(\int_{0}^{\mu+\frac{d+1}{L}}\rmd u\,f(u)\Big)+\sum_{r=1}^{\infty}\frac{B_{r}}{r!L^{r-1}}f^{(r-1)}\Big(\mu+\frac{d+1}{L}\Big)\nonumber\\
\hspace{68mm}
+\sum_{k=0}^{\infty}\frac{f_{k}}{L^{k/2}}\,\zeta(-\tfrac{k}{2},\,d+1)\;,
\end{eqnarray}
with coefficients $f_{k}$ given by (\ref{f[f]}). The sum over $r$ in the previous expression is the usual Euler-Maclaurin formula corresponding to the upper bound $j=\mu L$ of the sum (\ref{S(L,M)}). The sum over $k$ corresponds to the lower bound $j=1$ of the sum (\ref{S(L,M)}): there, one can not use the usual Euler-Maclaurin formula since the derivatives $f^{(r-1)}(0)$ are infinite. One can in fact write the sum over $r$ in the previous equation in a similar form as the sum over $k$:
\begin{equation}
\int_{\mu}^{\mu+\frac{a}{L}}\rmd u\,f(u)+\sum_{r=1}^{\infty}\frac{B_{r}\,f^{(r-1)}(\mu+\frac{a}{L})}{r!\,L^{r}}=-\sum_{r=1}^{\infty}\frac{f^{(r-1)}(\mu)\,\zeta(1-r,a)}{(r-1)!\,L^{r}}\;,
\end{equation}
which is a consequence of the identity for $r\in\mathbb{N}^{*}$
\begin{equation}
\label{Br(z)}
B_{r}(z)=\sum_{k=0}^{r}\C{r}{k}B_{r-k}\,z^{k}=-r\zeta(1-r,z)\;,
\end{equation}
where $B_{r}(z)$ is the $r$-th Bernoulli polynomial.

We also treat the case of a function $f$ with square root singularities at both ends $j=1$ and $j=N=\rho L$ of the sum (\ref{S(L,M)}):
\begin{equation}
\label{f[f,fbar]}
f(x)=\sum_{k=0}^{\infty}f_{k}\,x^{k/2}=\sum_{k=0}^{\infty}\bar{f}_{k}\,(\rho-x)^{k/2}\;.
\end{equation}
This is the case used in section \ref{section asymptotic expansions}. We assume that the first expansion is valid for $0\leq\arg x\leq\arg(d+1)$ and the second for $\arg(-d)\leq\arg(\rho-x)\leq0$ (if $d$ is in the upper half of $\mathbb{C}$; otherwise, one can replace $f(x)$ by $f(\rho-x)$ and $d$ by $-d-1$). Then, one can introduce an intermediate point $M=\mu L$, $0<\mu<\rho$ and write
\begin{equation}
\sum_{j=1}^{N}f\Big(\frac{j+d}{L}\Big)=\sum_{j=1}^{M}f\Big(\frac{j+d}{L}\Big)+\sum_{j=1}^{N-M}f\Big(\rho-
\frac{j-d-1}{L}\Big)\;.
\end{equation}
Using the relation
\begin{equation}
\label{Br(1-z)}
B_{r}(1-z)=(-1)^{r}B_{r}(z)\;
\end{equation}
on the Bernoulli polynomials, we observe that all the Bernoulli numbers from (\ref{S[B,zeta]}) cancel. One finds for $\rho>0$ the asymptotic expansion
\begin{eqnarray}
\label{S[zeta]}
\begin{picture}(0,0)(2,-8.5)
  \put(0,0){\line(1,0){107}}
  \put(0,-29){\line(1,0){107}}
  \put(0,0){\line(0,-1){29}}
  \put(107,0){\line(0,-1){29}}
\end{picture}
\sum_{j=1}^{\rho L}f\Big(\frac{j+d}{L}\Big)
\simeq L\Big(\int_{0}^{\rho}\rmd u\,f(u)\Big)+\sum_{k=0}^{\infty}\frac{f_{k}}{L^{k/2}}\,\zeta(-\tfrac{k}{2},\,d+1)\\
\hspace{60mm}
+\sum_{k=0}^{\infty}\frac{\bar{f}_{k}}{L^{k/2}}\,\zeta(-\tfrac{k}{2},\,-d)\;,\nonumber
\end{eqnarray}
with coefficients $f_{k}$ and $\bar{f}_{k}$ given by (\ref{f[f,fbar]}) and $\zeta$ the Hurwitz zeta function (\ref{zeta def}).
\end{section}

\section*{References}

%\bibliographystyle{unsrt}
%\bibliography{/users/prolhac/bib/references.bib}
%\bibliography{references}

\end{document}